\documentclass[10pt,journal,cspaper,compsoc]{IEEEtran} 
\usepackage{geometry}
\usepackage{filecontents}
\usepackage{amsmath,amssymb,amsfonts}
\usepackage{amsthm} 
\usepackage{color, colortbl}
\usepackage{graphicx}
\graphicspath{{fig/}}
\usepackage{xspace,xcolor,amsmath,mathtools,url,amsfonts,amssymb}
\usepackage{extarrows}
\usepackage{threeparttable}
\usepackage{arydshln}
\usepackage{booktabs}
\usepackage{subfig}
\usepackage[font=small,labelfont=bf]{caption}

\usepackage{setspace}

\usepackage{float}
\usepackage{subfloat}
\usepackage{setspace}
\usepackage{adjustbox}
\usepackage{textcomp}
\usepackage{xcolor}
\usepackage{hyperref}
\usepackage{multirow}
\usepackage{color}
\usepackage{protocolj,arydshln}
\usepackage{bm}
\usepackage{enumerate} 
\usepackage[inline,shortlabels]{enumitem}
\usepackage{xspace}
\usepackage{tikz}
\usepackage{amsmath}
\usepackage{adjustbox}
\newcolumntype{R}[2]{%
    >{\adjustbox{angle=#1,lap=\width-(#2)}\bgroup}%
    l%
    <{\egroup}%
}

\usepackage[libertine,liby,libaltvw,vvarbb]{newtxmath}

\newcommand*{\bigepsilon}{\mbox{\large$\epsilon$}}

\usepackage{pifont}
\newcommand{\cmark}{\ding{51}}%
\newcommand{\xmark}{\ding{55}}%

\usepackage{amsmath}
\usepackage{lipsum}
\usepackage{etoolbox}
\newcommand{\zerodisplayskips}{%
  \setlength{\abovedisplayskip}{1pt}%
  \setlength{\belowdisplayskip}{1pt}%
  \setlength{\abovedisplayshortskip}{1pt}%
  \setlength{\belowdisplayshortskip}{1pt}}
\appto{\normalsize}{\zerodisplayskips}
\appto{\small}{\zerodisplayskips}
\appto{\footnotesize}{\zerodisplayskips}

\newcommand{\secp}{\lambda}
\newcommand{\msg}{m\xspace}

\newcommand{\Gen}{\mathrm{Gen}}

\newcommand{\Eval}{\mathrm{Eval}}

\newcommand{\VRF}{\ensuremath{\mathrm{VRF}}\xspace}

\newcommand{\wits}{\ensuremath{w}\xspace}
\newcommand{\stmt}{\ensuremath{y}\xspace}
\newcommand{\resp}{\ensuremath{z}\xspace}

\newcommand{\chlgc}{\ensuremath{c}\xspace}

\newcommand{\strwits}{\textsf{wits}\xspace}
\newcommand{\strstmt}{\textsf{stmt}\xspace}
\newcommand{\strchlg}{\textsf{chlg}\xspace}
\newcommand{\strresp}{\textsf{resp}\xspace}

\newcommand{\zkproof}{\ensuremath{\pi}}
\newcommand{\Prove}{\mathrm{Prove}}
\newcommand{\Verify}{\mathrm{Verify}}

\newcommand{\st}{\ensuremath{\mathrm{st}}}

\newcommand{\rand}{\ensuremath{r}}

\newcommand{\NIZK}{\textrm{NIZK}\xspace}

\newcommand{\OWF}{\ensuremath{\mathrm{OWF}}\xspace}

\newcommand{\PRG}{\ensuremath{\mathrm{PRG}}\xspace}

\newcommand{\polyf}{\ensuremath{\mathrm{f}}\xspace}

\newcommand{\coeff}{\ensuremath{\alpha}\xspace}
\newcommand{\ssk}{\ensuremath{\mathrm{ssk}}\xspace}

\newcommand{\cmt}{\ensuremath{\mathrm{cmt}}\xspace}
\newcommand{\vvk}{\ensuremath{\mathrm{vvk}}\xspace}
\newcommand{\revsk}{\ensuremath{\overline{\mathrm{vsk}}}\xspace}

\newcommand{\hashh}{\ensuremath{h}\xspace}

\newcommand{\blockB}{\ensuremath{\mathrm{B}}\xspace}

\newcommand{\targetT}{\textsc{T}}

\newcommand{\strround}{\ensuremath{\mathrm{round}}\xspace}

\newcommand{\Fishsrt}{\ensuremath{\mathrm{s}}\xspace}
\newcommand{\Fishpk}{\ensuremath{\mathrm{y}}\xspace}

\newcommand{\signkey}{\mathrm{key}\xspace}

\newcommand{\rkey}{\ensuremath{\mathrm{k}}\xspace}

\newcommand{\slot}{\ensuremath{\mathrm{slot}}\xspace}

\newcommand{\msgm}{\ensuremath{\mathrm{m}}\xspace}

\newcommand{\msgx}{\ensuremath{\mathrm{x}}\xspace}

\newcommand{\vrfchlg}{\ensuremath{\mathrm{c}}\xspace}
\newcommand{\vrfmsg}{\ensuremath{\mathrm{m}}\xspace}

\newcommand{\seed}{\ensuremath{sd}\xspace}

\newcommand{\data}{\ensuremath{\mathrm{dt}}\xspace}

\newcommand{\uSign}{\ensuremath{\mathrm{uSign}}\xspace}

\newcommand{\ZKBpp}{\ensuremath{\mathrm{ZKB}\text{++}}\xspace}
\newcommand{\ZKBoo}{\ensuremath{\mathrm{ZKBoo}}\xspace}
\newcommand{\ZKB}{\ensuremath{\ZKBoo}\xspace}
\newcommand{\zksnark}{\ensuremath{\mathrm{ZK}\text{-}\mathrm{SNARK}}\xspace}
\newcommand{\zkstark}{\ensuremath{\mathrm{ZK}\text{-}\mathrm{STARK}}\xspace}

\newcommand{\Oursign}{\ensuremath{\sigma}\xspace}
\newcommand{\Oursk}{\ensuremath{s}\xspace}
\newcommand{\Ourpk}{\ensuremath{\alpha}\xspace}

\newcommand{\opstr}{\ensuremath{\mathrm{os}}\xspace}
\newcommand{\stt}{\ensuremath{\mathrm{st}}\xspace}

\newcommand{\DVRF}{\ensuremath{\mathrm{DVRF}}\xspace}
\newcommand{\DistKG}{\ensuremath{\mathrm{DistKG}}\xspace}
\newcommand{\PEval}{\ensuremath{\mathrm{PartialEval}}\xspace}
\newcommand{\Comb}{\ensuremath{\mathrm{Combine}}\xspace}

\usepackage{cancel}
\usepackage[normalem]{ulem}
\newcommand{\draft}[1]{}
\newcommand{\ignore}[1]{}
\providecommand{\internalFlag}{0}
\ifnum\internalFlag=1
\newcommand{\bigtodo}[1]{}
\newcommand{\todo}[1]{}

\newcommand{\fnote}[1]{}
\newcommand{\hnote}[1]{}
\newcommand{\hsnote}[1]{}
\newcommand{\zli}[1]{}
\newcommand{\zpnote}[1]{}
\newcommand{\linote}[1]{}
\newcommand{\zlinote}[1]{}
\newcommand{\zlimark}[1]{}
\newcommand{\zliins}[1]{}
\newcommand{\zlidel}[1]{}

\newcommand{\zlirevise}[1]{}
\newcommand{\zlidelete}[1]{}

\else

\newcommand{\fnote}[1]{{\color{red}{\bf [Lei: #1]}}}

\newcommand{\hsnote}[1]{{\color{blue}{\bf [{\bf Hongsheng:} #1]}}}
\newcommand{\hnote}[1]{{\color{blue}{\bf [{\bf Hongsheng:} #1]}}}

\newcommand{\zli}[1]{{\color{blue}{\footnotesize [{\rm ZP:} #1]}}}
\newcommand{\zpnote}[1]{{\color{red}{\footnotesize [{\bf ZP:} #1]}}}
\newcommand{\linote}[1]{{\color{red}{\footnotesize[{ZP:} #1]}}}
\newcommand{\zlinote}[1]{{\color{red}{\footnotesize [{ZP:} #1]}}}
\newcommand{\zlimark}[1]{{\color{magenta}{[ #1]}}\xspace}

\definecolor{lgray}{gray}{0.5}

\newcommand{\zliins}[1]{\small{{\color{cyan}{\footnotesize [Insert:] #1}}}}
 \newcommand{\zlidel}[1]{\small{{\color{lgray}{\rm [Delete:]#1}}}}

\newcommand{\todo}[1]{%
    \fcolorbox{red!50!black}{red!20}{%
        #1%
    }
}

\newcommand{\bigtodo}[1]{%
    \begin{center}
    \fcolorbox{red!50!black}{red!20}{%
        \parbox{0.8\linewidth}{#1}%
    }
    \end{center}
}

\fi

\newcommand{\ie}{\textit{i.e.}\xspace}

\newcommand{\eg}{\textit{e.g.}\xspace}
\newcommand{\etc}{\textit{etc}\xspace}

\newcommand{\etal}{\textit{et~al.}\xspace}

\newlength{\protowidth}

\newcommand{\mybox}[5]{
    \begin{figure}[!ht]
        \centering
    \begin{tikzpicture}
        \node[anchor=text,text width=\columnwidth-0.3cm, draw, rounded corners, line width=1pt, fill=#3, inner sep=1.8mm] (big) {\\#4};
        \node[draw, rounded corners, line width=.5pt, fill=#2, anchor=west, xshift=5mm] (small) at (big.north west) {#1};
    \end{tikzpicture}
    \caption{#5}
    \end{figure}
    \vspace{1mm}
}

\usepackage{tcolorbox}
\usepackage{fancybox}
\usepackage[framemethod=TikZ]{mdframed}
\definecolor{mycolor}{rgb}{0.122, 0.435, 0.698}
\newmdenv[innerlinewidth=0.5pt, roundcorner=4pt,linecolor=mycolor,innerleftmargin=6pt,
innerrightmargin=6pt,innertopmargin=6pt,innerbottommargin=6pt]{zlibox}
\newcommand{\aaa}{\ensuremath{\mathrm{a}}\xspace}

\newcommand{\ccc}{\ensuremath{\mathrm{c}}\xspace}

\newcommand{\zzz}{\ensuremath{\mathrm{z}}\xspace}

\newcommand{\G}{\ensuremath{\mathbb{G}}}

\newcommand{\R}{\ensuremath{\mathbb{R}}}
\newcommand{\Z}{\ensuremath{\mathbb{Z}}}

\newcommand{\Zq}{\ensuremath{\Z_q}}

 \newtheorem{theorem}{Theorem}[section]

 \newtheorem{remark}[theorem]{Remark}
 
\numberwithin{equation}{section}

\renewcommand{\epsilon}{\varepsilon}

\DeclareMathOperator{\poly}{poly}
\DeclareMathOperator{\polylog}{polylog}
\DeclareMathOperator{\negl}{negl}

\newcommand{\calD}{\ensuremath{\mathcal{D}}}

\newcommand{\calK}{\ensuremath{\mathcal{K}}}

\newcommand{\calL}{\ensuremath{\mathcal{L}}}

\newcommand{\calR}{\ensuremath{\mathcal{R}}}

\DeclarePairedDelimiter\myset{\{}{\}}

\newcommand{\key}[1]{\ensuremath{#1}}
\newcommand{\pk}{\key{pk}}
\newcommand{\vk}{\key{vk}}
\newcommand{\sk}{\key{sk}}

\newcommand{\attacker}[1]{\ensuremath{\mathcal{#1}}}

\newcommand{\Com}{\mathrm{Com}}
\newcommand{\Open}{\mathrm{Open}}

\newcommand{\VrfGen}{\ensuremath{\mathrm{VrfGen}}\xspace}

\newcommand{\VrfEval}{\ensuremath{\mathrm{VrfEval}}\xspace}

\newcommand{\VrfVrfy}{\ensuremath{\mathrm{VrfVrfy}}\xspace}

\newcommand{\vpk}{\ensuremath{\mathrm{vpk}}\xspace}
\newcommand{\vsk}{\ensuremath{\mathrm{vsk}}\xspace}
\newcommand{\vrfpf}{\ensuremath{{\pi_{\mathrm{vrf}}}}\xspace}
\newcommand{\vrfout}{\ensuremath{{\mathrm{out}_{\mathrm{vrf}}}}\xspace}

\renewcommand{\to}{\ensuremath{\rightarrow}}

\newcommand{\abs}[1]{\ensuremath{\lvert{#1}\rvert}}

\newcommand{\set}[1]{\ensuremath{\left\{#1\right\}}}

\newcommand{\state}{\ensuremath{\mathrm{st}}\xspace}

\newcommand{\Sign}{\ensuremath{\mathrm{Sign}}\xspace}
\newcommand{\Vrfy}{\ensuremath{\mathrm{Vrfy}}\xspace}

\newcommand{\hashF}{\ensuremath{\mathrm{F}}\xspace}
\newcommand{\hashH}{\ensuremath{H}\xspace}
\newcommand{\hashG}{\ensuremath{\mathrm{G}}\xspace}

\usetikzlibrary{arrows,positioning, shapes, calc, backgrounds, fit}
\usetikzlibrary[shapes.arrows]
\usetikzlibrary{shapes.geometric}
\usetikzlibrary{arrows}
\usetikzlibrary{backgrounds}
\usetikzlibrary{matrix}
\usetikzlibrary{positioning}
\usetikzlibrary{calc}
\usetikzlibrary{intersections}
\usetikzlibrary{fadings}
\usetikzlibrary{decorations.footprints}
\usetikzlibrary{decorations.pathmorphing}
\usetikzlibrary{decorations.pathreplacing}
\usetikzlibrary{patterns}
\usetikzlibrary{shapes.callouts}
\usetikzlibrary{fit}
\usetikzlibrary{arrows,positioning, shapes, calc, backgrounds, fit, shadows}
\usetikzlibrary{matrix,shapes,arrows,positioning,chains,calc}
\usetikzlibrary[shapes.arrows]
\usetikzlibrary{shapes.geometric}
\usetikzlibrary{arrows}
\usetikzlibrary{backgrounds}
\usetikzlibrary{matrix}
\usetikzlibrary{positioning}
\usetikzlibrary{calc}
\usetikzlibrary{intersections}
\usetikzlibrary{fadings}
\usetikzlibrary{decorations.footprints}
\usetikzlibrary{decorations.pathmorphing}
\usetikzlibrary{decorations.pathreplacing}
\usetikzlibrary{patterns}
\usetikzlibrary{shapes.callouts}
\usetikzlibrary{fit}
\usetikzlibrary{matrix}
\usetikzlibrary{calc}
\usetikzlibrary{shadows}
\usetikzlibrary{scopes}
\usetikzlibrary{chains}
\usetikzlibrary{positioning}
\usetikzlibrary{fit}
\usetikzlibrary{backgrounds}
\usetikzlibrary{decorations.pathmorphing}
\usetikzlibrary{decorations.pathreplacing}
\usetikzlibrary{shapes.geometric}
\usetikzlibrary{shapes.symbols}
\usetikzlibrary{shapes.misc}
\usetikzlibrary{arrows}
\usetikzlibrary{shapes,arrows}
\usetikzlibrary{shapes.symbols,patterns}
\usetikzlibrary{shapes,positioning,decorations.pathmorphing}
\tikzset{
  every shadow/.style={opacity=.4,shadow xshift=.3ex,shadow yshift=-.3ex},
  every matrix/.style={ampersand replacement=\&},
  every edge/.append style={->,thick},
  every join/.style={->,thick},
}

\tikzset{
  Ztrans/.style={x={(1.5cm,1cm)},y={(.5cm,1.5cm)}},
  Zdual/.style={x={(1.5cm,-.5cm)},y={(.5cm,1cm)}},
  latttrans/.style={x={(1.8cm,.5cm)},y={(0.4cm,1.3cm)}},
  dualtrans/.style={x={(1.3cm,-.4cm)},y={(-.5cm,1.8cm)}},
  lattB/.style={x={(1.8cm,.5cm)},y={(0.4cm,1.4cm)}},
  dualB/.style={x={(1.4cm,-.4cm)},y={(-.5cm,1.8cm)}},
  shortB/.style={x={(.5cm,.2cm)},y={(-.1cm,1.7cm)}},
  latt2B/.style={x={(1cm,.1cm)},y={(.1cm,1.2cm)}},
  ulattB/.style={x={(2.0cm,.5cm)},y={(2.1cm,-.2cm)}},
  openball/.style={fill=Blue!20},
  ball/.style={openball,thin,draw=Blue},
  latt/.style={Lattice,fill},
  dist/.style={Brown,densely dashed,font=\footnotesize},
  offlabel/.style={Brown,inner sep=2pt,font=\footnotesize},
  axes/.style={->,Gray,style=very thin},
  fundamental/.style={Gray,fill opacity=.2},
  target/.style={fill,Brown},
  project/.style={thin,dashed},
  help lines/.style={Gray,very thin,step=.5cm},
  noise/.style={
    Lattice,
    opacity=.5,
    decorate,decoration={
      expanding waves,
      segment length=5pt,angle=10,
      post=moveto,post length=3pt
    },
  },
}

\tikzset{
    ncbar angle/.initial=90,
    ncbar/.style={
        to path=(\tikztostart)
        -- ($(\tikztostart)!#1!\pgfkeysvalueof{/tikz/ncbar angle}:(\tikztotarget)$)
        -- ($(\tikztotarget)!($(\tikztostart)!#1!\pgfkeysvalueof{/tikz/ncbar angle}:(\tikztotarget)$)!\pgfkeysvalueof{/tikz/ncbar angle}:(\tikztostart)$)
        -- (\tikztotarget)
    },
    ncbar/.default=0.5cm,
}

\tikzset{square left brace/.style={ncbar=0.5cm}}
\tikzset{square right brace/.style={ncbar=-0.5cm}}

\tikzset{round left paren/.style={ncbar=0.5cm,out=120,in=-120}}
\tikzset{round right paren/.style={ncbar=0.5cm,out=60,in=-60}}

\tikzstyle{vecArrow} = [thick, decoration={markings,mark=at position
   1 with {\arrow[semithick]{open triangle 90}}},
   double distance=1.5pt, shorten >= 4.5pt,
   preaction = {decorate},
   postaction = {draw,line width=2pt, white,shorten >= 4.5pt}]

\usetikzlibrary{calc,shapes.callouts}

\definecolor{gauss}{rgb}{0.8,0.2,0.2}
\definecolor{RegionColor}{rgb}{0.89,0.89,0.58}
\definecolor{Math}{rgb}{0.05,0.05,0.55}
\definecolor{Curva}{rgb}{0.10,0.20,0.30}
\definecolor{FillBox}{rgb}{1,0.8,0.2}
\definecolor{Facts}{rgb}{0.05,0.95,0.05}
\definecolor{Sign}{rgb}{0.7,0.95,0.7}
\definecolor{Ver}{rgb}{0.7,0.95,0.2}
\definecolor{pink}{rgb}{1,0.75,0.80}
\definecolor{v1}{rgb}{1,0.4,0.2}
\definecolor{v2}{rgb}{0.2,0.4,1}
\definecolor{pc}{rgb}{0.94,0.98,0.75}

\tikzstyle{Box}=[rectangle,draw=Curva,fill=FillBox,rounded corners, minimum width=1.3cm]
\tikzstyle{aBox}=[rectangle,draw=black,rounded corners, yshift=0.5cm]
\tikzstyle{acBox}=[rectangle,draw=black, fill=Sign, rounded corners, yshift=0.5cm, text width=1.3cm, align=center]
\tikzstyle{lBox}=[rectangle,draw=Curva,fill=FillBox,rounded corners, minimum width=1.3cm]
\tikzstyle{fBox}=[rectangle,draw=black, fill=Facts,rounded corners, minimum width=3cm]
\tikzstyle{mBox}=[rectangle,draw=black,rounded corners]
\tikzstyle{sBox}=[rectangle,draw=black]
\tikzstyle{pBox}=[rectangle,draw=black, fill=pc]
\tikzstyle{rBox}=[rectangle,draw=black, fill=red!95,rounded corners]
\tikzstyle{gBox}=[rectangle,draw=black, fill=green!75,rounded corners]
\tikzstyle{bBox}=[rectangle,draw=black, fill=blue!25,rounded corners]
\tikzstyle{Circ}=[circle,draw=black, fill=RegionColor!75]

\NewEnviron{elaboration}{
\par
\begin{tikzpicture}
\node[rectangle,minimum width=0.24\textwidth] (m) {\begin{minipage}{0.24\textwidth}\BODY\end{minipage}};
\draw[dashed] (m.south west) rectangle (m.north east);
\end{tikzpicture}
}

\usetikzlibrary{decorations.pathreplacing}
\definecolor{myLightGray}{RGB}{191,191,191}
\definecolor{myGray}{RGB}{160,160,160}
\definecolor{myDarkGray}{RGB}{144,144,144}
\definecolor{myDarkRed}{RGB}{167,114,115}
\definecolor{myRed}{RGB}{255,58,70}
\definecolor{myGreen}{RGB}{0,255,71} 

\newcommand{\ZKP}{\mathrm{ZKP}}
\newcommand{\word}{y}
\newcommand{\tape}{\mathsf{tp}}
\newcommand{\tapei}{\mathsf{tp}^{(i)}}

\newcommand{\zkbview}{\mathsf{view}}
\newcommand{\zkbviewi}{\mathsf{view}^{(i)}}
\newcommand{\zkbviewxx}[1]{\mathsf{view}^{(#1)}}
\newcommand{\zkbshare}{{y}}
\newcommand{\zkbsharei}{{y}^{(i)}}
\newcommand{\zkbwitsi}{\wits^{(i)}}
\newcommand{\Share}{\mathsf{Share}}
\newcommand{\Updatec}{\mathsf{Upd}_{\phi}}
\newcommand{\UpdateF}{\mathsf{Upd}_{\hashF}}
\newcommand{\UpdateH}{\mathsf{Upd}_{\hashH_2}}

\newcommand{\Output}{\mathsf{Output}}
\newcommand{\Reconst}{\mathsf{Rec}}
\newcommand{\ccti}{\mathsf{ct}^{(i)}}
\newcommand{\copi}{\mathsf{op}^{(i)}}

\newcommand{\cctip}{\mathsf{ct}'^{(i)}}

\newcommand{\mstmti}{a^{(i)}}
\newcommand{\mstmtxx}[1]{a^{(#1)}}
\newcommand{\mstmtxxp}[1]{a'^{(#1)}}
\newcommand{\wire}{\mathsf{w}}
\newcommand{\gate}{\mathsf{g}}

\begin{document}
\title{Post-Quantum VRF and its Applications in Future-Proof Blockchain System}
\author{Zengpeng~Li, 
        Teik Guan Tan,
        Pawel Szalachowski,
        Vishal Sharma,
         and~Jianying~Zhou
\IEEEcompsocitemizethanks{
\IEEEcompsocthanksitem Z.~Li is with the Shandong University Qingdao Campus, China. (E-mail: zengpeng@email.sdu.edu.cn.) 
\IEEEcompsocthanksitem T.G. Tan, P. Szalachowski and J. Zhou are with the Singapore University of Technology and Design, Singapore. (E-mail:~\{pawel, teikguan\_tan, jianying\_zhou\}@sutd.edu.sg.)
\IEEEcompsocthanksitem V. Sharma is with the Queen's University Belfast, UK, (E-mail:~v.sharma@qub.ac.uk.)}
}


\IEEEcompsoctitleabstractindextext{
\begin{abstract}
%
A verifiable random function (VRF in short) is a powerful pseudo-random function that provides a non-interactively public verifiable proof for the correctness of its output. Recently, VRFs have found essential applications in blockchain design, such as random beacons and proof-of-stake consensus protocols. To our knowledge, 
the first generation of blockchain systems used inherently inefficient proof-of-work consensuses, and the research community tried to achieve the same properties by proposing proof-of-stake schemes where resource-intensive proof-of-work is emulated by cryptographic constructions. 
Unfortunately, those most discussed proof-of-stake consensuses (\eg, Algorand and Ouroborous family) are not future-proof because the building blocks are secure only under the classical hard assumptions; in particular, their designs ignore the advent of quantum computing and its implications. 
In this paper, we propose a generic compiler to obtain the post-quantum VRF from the simple VRF solution using symmetric-key primitives (\eg, non-interactive zero-knowledge system) with an intrinsic property of quantum-secure. Our novel solution is realized via two efficient zero-knowledge systems \ZKB and \ZKBpp, respectively, to validate the compiler correctness. Our proof-of-concept implementation indicates that even today, the overheads introduced by our solution are acceptable in real-world deployments. We also demonstrate potential applications of a quantum-secure VRF, such as quantum-secure decentralized random beacon and lottery-based proof of stake consensus blockchain protocol.

\end{abstract}
}

\maketitle 
\IEEEdisplaynontitleabstractindextext
\IEEEpeerreviewmaketitle

\section{Introduction}\label{sec:intro}

Public distributed ledgers constitute a new class of systems. Recently they gained a lot of attention from academic and industry researchers, various businesses, governments, and other organizations, \eg,~\cite{TDSC:Wang19,TIFS:Li20}.  Thanks to distributed and authenticated append-only data structures, these systems provide transparency, availability, and censorship resistance. However, the main novelty and breakthrough introduced by public distributed ledgers is a consensus protocol that is open to anyone and does not require any privileged entities.
Currently, a lot of attention is attracted by proof of stake (PoS) blockchains, which solve
the energy inefficiency of proof of work (PoW) protocols.  Instead of holding a PoW-based competition to decide who is the round leader (adding a block of transactions), in PoS systems, a leader is chosen via a cryptographic lottery from available network nodes, with their chances of winning is proportional to the number of coins they hold. 

A worrying trend in the blockchain community is to ignore security lessons learned from the previous designs and deployments of critical systems.  It especially concerns as if adopted, and blockchains are supposed to ``provide trust'' and be long-lived. Thus their security is critical.  Besides common security threats, some potential risks can be foreseen.  One of the most severe is the advent of quantum computing as a functional quantum computer could easily undermine the security of the most distributed ledgers, making
them practically useless. 
Although,
quantum computing is still a premature organization, like NIST and Google, they
concern about post-quantum cryptographic methods to mitigate risk for quantum computer destruction~\cite{chen2016report}.  We do not see such a trend in the blockchain community, however. 

Additionally, many consensus protocols involve allocating the creation of block creator, whose selection procedure, most often than not, requires a method for collective randomness sampling. To our knowledge, computers are based on a Turing machine that is a deterministic device, and the same input seed always produces the same output sequence. Thus, computers are bad at the generation of randomnesses, and their outputs are pseudo-random. 
During the randomness sampling, adversarially biased hashes (known as grinding attacks or randomness-biasing attacks) imply that a (malicious) miner can re-create a block multiple times until it is likely that the miner can create a second block shortly afterward. 
In other words, the adversary can bias the nonce that is used to seed the hash since the adversary can place arbitrary seeds in the blocks it contributes.

Indeed, randomness-biasing (or grinding) attacks can be mitigated (even thwarted) by ensuring that a miner is not able to influence the next leader election
by using an unbiased source of randomness or a deterministic leader election. Obliviously, real-world entropy is not suitable for use as a seed for blockchain randomness~\cite{ACNS:CascudoD17,SP:SJKGGKFF17}. 
In reality, there are two main approaches to blockchain randomness in production, one is the novel approach of ``\textit{two-lookback}'' mechanism proposed by Snow White~\cite{FC:DaiPasShi19}, another is using the verifiable random function (\VRF) such as Algorand~\cite{Algorand16} and Ouroboros Praos~\cite{EC:DGKR18}, \etc,
or using RANDAO supported by verifiable delay function (VDF). 
%
Concretely, the randomness (\ie, seed) published at round $r$ is determined using a deterministic hashing function with the preceding randomness from $2\secp$ to $\secp$ in Snow White. Similarly, in Algorand,  the randomness (\ie, seed) published at round $r$ is determined using \VRF with the seed of the previous round $r-1$, \eg, $\seed_{r}=\hashH(\seed_{r-1}\| r)$, and the initial value of the seed $\seed_0$.
The uniqueness of the pseudorandom output of \VRF enhances the bias resistance, as it stands against any adversary independently from the number of corrupted servers that the adversary controls. However, Snow White only achieves a weak uniqueness with some tolerable errors, supported by strict requirements for the participants. \textit{An important observation is that these mentioned $\VRF$s are based on non-quantum-safe hardness assumptions so that they can not remain secure in the quantum computing era.} Thus, the observations as mentioned above promote us to ask the following question.
\begin{quote}
\textit{
Is it possible to propose a post-quantum verifiable random function that benefits random beacon and proof of stake consensus protocol in the coming quantum era?
}
\end{quote}

\par\noindent\textbf{Contributions and Technical Route}.
This question can be answered in the affirmative. In other words, our goal is to explore how to remain secure for the PoS blockchain consensus in the quantum era. Inspired by recent PoS-based blockchain systems~\cite{Algorand16,EC:DGKR18} and quantum-resistant cryptographic primitives ~\cite{CCS:CDGORR17,CCS:KatKolWan18}, the crux of fulfilling the main goal is turning into how to construct an efficient post-quantum \VRF based on the existing quantum-secure cryptographic primitives, \eg, symmetric-based cryptographic building blocks. The main reason is that the uniqueness of the pseudorandom output of \VRF enhances the bias resistance and benefits cryptographic lottery-based proof of stake consensus protocol.


Very recently, Kiltz~\etal~\cite{EC:KilLyuSch18} have provided deterministic signatures that are secure under the standard chosen message attack in the quantum random oracle model,
which is obtained via Fiat-Shamir transformation, a paradigm of
combining a hash function and an identification scheme to produce a
digital signature scheme.  Kiltz~\etal~\cite{EC:KilLyuSch18} also provide a
practical instantiation of a signature scheme, Dilithium-QROM, with a
tight security reduction in the QROM from the Module-LWE assumption over the lattice.
Although various lattice-based pseudorandom functions (in short PRFs) are proposed in succession, these PRFs are not enabling to achieve the \textit{public verifiability} and guarantee to play the role of the traditional $\hashH(\cdot)$ in the large-scale network due to the overhead they introduce.

Below, we summarize our main contributions along with the technical route.

\begin{itemize}

\item \textbf{Post-Quantum VRF}. We construct a post-quantum \VRF  from symmetric-key primitives by using the post-quantum ZKP systems in \autoref{sec:framework}. In a nutshell, the methodology is using a pseudo-random function to generate the output of \VRF $\Oursign \gets  \hashH_2(\vsk, \hashh)$ for  $\hashh=\hashH_1(\msgm)$ with the secret key $\vsk$ while introducing a quantum-safe zero-knowledge proof (ZKP) system to append a proof of knowledge of secret key $\vsk$, where the proof is used to show that $\hashh$ is pre-image of the output $\Oursign$  (without revealing $\vsk$). Thus, we effectively have a deterministic output with \textit{uniqueness} functionality.

\item \textbf{Post-Quantum Random Beacon via proposed VRF}. To prevent the adversary from predicting who is the following block creator while remaining secure in the quantum era, we offer a quantum-safe decentralized random beacon without depending on a third party. 
In that case, a straightforward way,
either using the ``two-lookback'' mechanism to hash past $\secp$ blocks in Snow White or using \VRF to hash the previous block in Algorand and Ouroboros Paros, \etc, to generate randomness is not a proper approach. Thus, post-quantum decentralized randomness is generated using the technique of quantum-safe distributed-\VRF  in \autoref{sec:beacon}, which is achieved by combining the techniques of verifiable secret sharing and \VRF from symmetric primitives.

\item \textbf{Post-Quantum PoS via proposed VRF}. $\VRF$ is becoming the core of the cryptographic lottery-based PoS consensus protocol. The crux of fulfilling the goal of post-quantum PoS is turning into the instantiation of post-quantum \VRF. In that case, armed with the constructed post-quantum \VRF, as discussed in \autoref{sec:VRF}, we enable to eliminate the \VRF based on number theoretical assumptions in Algorand, and we only require that in each round each node evaluates a post-quantum \VRF to check whether they have been sampled as part of that round's committee. Meanwhile, the number of committee members is binomial (\ie, depending on the stake distribution), so the amount of work needed to verify the other committees' messages is also binomial.

\end{itemize}

%
%
%
%
%
%
%

\noindent\textbf{Organization}.
This work is organized as follows.
The next~\autoref{sec:Preliminaries} is for preliminaries. 
In \autoref{sec:framework} we present
our motivations and observations, and we discuss how to construct a post-quantum \VRF with an implementation and evaluation of our concrete instantiation.
In \autoref{sec:beacon}, we present the first application of \VRF in random beacon. 
In \autoref{sec:VRF}, then we present the second application of \VRF in proof of stake.
Finally, \autoref{sec:concl} is for conclusion.


\section{Related Work}\label{sec:related}
 The concept of $\VRF$ is very similar to the concept of signature. Thus,  there are two main approaches for the construction of VRFs: 1) a direct approach (\eg, \cite{NSEC5}), or 2) an indirect approach using signatures that have the ``uniqueness'' property (\eg, \cite{C:GolOst92}). There is also a  close relation between $\VRF$s, signatures, and $\ZKP$s. Below, we revisit some related works.


%

\subsection{Revisiting Verifiable Random Function}
To our knowledge, $\VRF$ is an important cryptographic tool in blockchain consensus and random beacon. 
The essence of \VRF is a mathematical operation that takes some inputs and produces a random number along with proof of authenticity that the random number was generated by the submitter. The appended proof can be verified by any challenger to ensure the random number generation is valid.  
To our knowledge, Goldwasser and Ostrovsky~\cite{C:GolOst92} introduced an invariant signature that is called a unique signature afterward, and they have shown how to transform a unique signature to the \VRF in a fairly straightforward way. Subsequently, Micali, Rabin, and Vadhan~\cite{FOCS:MicRabVad99} pointed out that if a digital signature scheme is with the uniqueness property, then ideal hashing and unique signature provide an elementary implementation of a \VRF. 
The existing constructions of \VRF~\cite{NSEC5,PKC:Kohl19} based on several theoretical assumptions cannot guarantee security in the coming quantum era. Recently, lattice-based \VRF schemes are proposed, \eg,~\cite{C:Yamada17}, but these schemes do not have good performance in the real world. 
Thus, achieving a practical post-quantum \VRF is an open question. Very recently, an independent and similar result with us was proposed by Buser \etal~\cite{EPRINT:BDEKKL21}, they used the XMSS signature as the building block to design quantum-secure VRF, their methodology is using the unique signature (\eg, XMSS signature) as the beginning point.

{
%
\definecolor{Gray}{gray}{0.9}
\definecolor{LightCyan}{rgb}{0.88,1,1}
\newcolumntype{a}{>{\columncolor{Gray}}c}
\newcolumntype{b}{>{\columncolor{white}}c}
\begin{table*}[ht!]
\vspace{-2pt}
\begin{center}
\begin{threeparttable}
    \caption{Overview of Feasibly Quantum-Resistant Proof Systems.}
    \label{table:pqnizk}
    \begin{tabular}{l a b a  b  a  b }
      \toprule 
      \multicolumn{1}{c}{\multirow{2}{*}{\textbf{Scheme}}}
      & \multicolumn{2}{c}{\underline{\textbf{Performance}}}
      & \multicolumn{2}{c}{\underline{\textbf{Size}}}
      & \multicolumn{1}{c}{\underline{\textbf{Assumption}}}
      & \multicolumn{1}{c}{\underline{\textbf{Untrusted}}}\\
      & \textbf{Prover} & \textbf{Verifier} & \textbf{CRS} & \textbf{Proof}  & $\sk$ & \textbf{Setup} \\
      [0.5ex]
      \midrule 
      \multirow{1}{*}{$\ZKB$~\cite{USENIX:GiaMadOrl16}}
      & $n\log{n}$ & $c\log{c}+h\log{h}$  & $n$ & $\sqrt{n}$ & CRHF & \xmark\\
      \hdashline
      \multirow{1}{*}{Ligero~\cite{CCS:AHIV17}}
      & $n\log{n}$ & $c\log{c}+h\log{h}$  & $0$ & $\sqrt{n}$ & CRHF & \cmark\\
      \hdashline
      \rowcolor[gray]{0.9}
      \multirow{1}{*}{$\zkstark$~\cite{EPRINT:BBHR18}}
      & $n\polylog{n}$ & $\polylog{n}$ & $0$ &  $\log^2{n}$ & CRHF & \cmark\\
      \hdashline
      \multirow{1}{*}{Aurora~\cite{EC:BCRSVW19}}
      & $n\log{n}$ & $c\log{c}+h\log{h}$  & $0$ & $\sqrt{n}$ & CRHF & \cmark\\
      \hdashline
      \multirow{1}{*}{Bootle~\etal~\cite{AC:BCGGHJ17}}
      & $n$ & $n$ & $0$ & $\sqrt{n}$ & CRHF & \cmark\\
      \bottomrule 
    \end{tabular}
    \begin{tablenotes}
  \item [-] The communication complexity of $\ZKBpp$ is more than halved compared to $\ZKB$, not affecting the computational complexity. This is done by using six different optimizations that are designed to compress all messages sent as much as possible.
  \item[-] $n$ is the number of gates, $d$ is the depth of the circuit, $h$ is the width of the subcircuits, $c$ is the number of copies of the subcircuits, $\ell$ is the size of the instance, and $w$ is the size of the witness.
  \item[-]  
  CRHF stands for collision-resistant hash functions.
  \item [-] \cmark~ denotes that the scheme does have this property.
  \item [-] \xmark~ denotes that the scheme does not have this property.
      \end{tablenotes}
\end{threeparttable}
\end{center}
\vspace{-2em}
\end{table*}
}

\subsection{Revisiting Post-Quantum Zero-Knowledge} 
Recent advances in quantum computing have increased the interest in post-quantum cryptography research, such as lattice-based, code-based, multivariate-quadratic (MQ), and hash-based cryptography. Lattice-based cryptographic is more versatile than others, it has a solid theoretical foundation and problems (\eg, CVP, SVP, SIS, and LWE) and can realize various cryptographic primitives. and hash-based cryptography enables to reduce greatly the size of the $\ZKP$' statement by using pseudorandom functions instead of signatures wherever possible.

Notably, there exist ZKPs that only use collision-resistant hash functions and hence are plausibly post-quantum secure. Popular examples include  Aurora~\cite{EC:BCRSVW19},
$\ZKB$~\cite{USENIX:GiaMadOrl16}, $\ZKBpp$~\cite{CCS:CDGORR17}, and $\zkstark$ (zero-knowledge scalable transparent argument of knowledge)~\cite{EPRINT:BBHR18}. 
Other notable ZKPs like Bulletproofs or libsnark rely on cryptographic assumptions that are known to be vulnerable to quantum attacks (\eg, the hardness of discrete log in certain elliptic curve groups). Inspired by the work of Sonic~\cite{EPRINT:MBKM19}, Table~\ref{table:pqnizk} is used to illustrate asymptotic efficiency comparison of the post-quantum $\ZKP$ security for arithmetic circuits.  
$\zkstark$~\cite{EPRINT:BBHR18} aims to provide fast and scalable solutions while ensuring financial security. Armed with $\zkstark$, transaction encryption is possible. 
The acronym $\zksnark$ means ``Zero-Knowledge Succinct Non-Interactive Argument of Knowledge," the most important property is that the proof length does not depend on the length of statements. 
There are currently two leading technologies (\ie, Monero, and ZCash) offering their cryptocurrencies while striving to solve protection issues. In particular,  Monero uses $\zkstark$ that provides the ability to conduct anonymous transactions. However, Zcash is supported by $\zksnark$ to create a shielded transaction.
Compared with  $\zksnark$,
Bulletproof is the short non-interactive zero-knowledge range proof that requires no trusted setup, while verifying a bulletproof is more time-consuming than verifying a SNARK proof.


%

%



{
\definecolor{Gray}{gray}{0.9}
\definecolor{LightCyan}{rgb}{0.88,1,1}
\newcolumntype{a}{>{\columncolor{Gray}}c}
\newcolumntype{b}{>{\columncolor{white}}c}
\begin{table}[h!]
\vspace{-3pt}
\scriptsize
\begin{center}
\begin{threeparttable}
    \caption{Overview of Post-Quantum Signature Schemes.}
    \label{table:pqsystem}
    \begin{tabular}{l l a  b  a }
      \toprule 
      \multicolumn{2}{c}{\multirow{2}{*}{\textbf{Performance}}} & \multicolumn{3}{c}{Size[Byte]}\\
      & &$\Sign$ & $\pk$  & $\sk$  \\
      [0.5ex]
      \midrule 
      \multirow{3}{*}{NTRU-based}
      & NTRU-GPV~\cite{AC:DucLyuPre14} & $\approx$ 1200 & 1792 &  NONE \\
      & BLISS-BI~\cite{C:DDLL13} & 717 & 896 & 256 \\
      & Falcon-512~\cite{FALCON17}$^{*}$ & 617 & 897 & 4097  \\
      \hdashline
      \multirow{2}{*}{Ring-based}
      & Dilithium-medium~\cite{TCHES:DKLLSSS18} & 2044 & 1184 &  2800 \\
      & Dilithium-recommended~\cite{TCHES:DKLLSSS18} & 2071 & 1472 & 3504 \\
      \hdashline
      \multirow{2}{*}{Lattice-based}
      & qTESLA-p-I~\cite{EPRINT:ABBLR19} & 2848 & 14880 & 5184 \\
      & qTESLA-p-III~\cite{EPRINT:ABBLR19} & 6176 & 39712 & 12352 \\
       \hdashline
      \multirow{3}{*}{MQ-based}
      & MQDSS-31-64~\cite{AC:CHRSS16} & 64  & 34032& 24 \\
      & HmFev(256,15,3,16)~\cite{IEICE:CLPYC18} & 61 &  83100& $\approx$20\\
      & Rainbow(16,32,32,32)~\cite{IEICE:CLPYC18} & 48 & 145500 & 18 \\
      \hdashline
      \multirow{3}{*}{Hash-based}
      & SPHINCS$^{+}-128$~\cite{EPRINT:ABBLR19} & 16976 & 32 & 64 \\
      & Picnic\_L1\_FS~\cite{EPRINT:ABBLR19} & 32944 & 33 & 49 \\
      & Picnic2\_L1\_FS~\cite{EPRINT:ABBLR19} & 12062 & 33 & 49 \\
      \bottomrule 
    \end{tabular}
    \begin{tablenotes}
        \footnotesize
  \item[*] {\href{https://github.com/open-quantum-safe/liboqs}{https://github.com/open-quantum-safe/liboqs}} 
      \end{tablenotes}
\end{threeparttable}
\end{center}
\vspace{-3em}
\end{table}
}

\subsection{Revisiting Post-Quantum Signatures}
Obviously, there are similarities between the \VRF and the unique signature, and some unique results enable to provide a \VRF in a straightforward way. According to our investigation~\cite{NSEC5}, there is a direct approach to construct $\VRF$, which can bypass the unique signature as intermediation. To achieve quantum-safe, 
%
%
No one doubts that lattice-based cryptography still objects to further research, but hash-based signatures are well understood.
There are many reasons to use hash-based schemes 
The interesting point is that the hash-based signature could reduce a message to a small representation of characters that can be signed easily. Whereas a hash-based solution only needs a secure hash function for the same procedure.


Below,  we revisit classical digital signature schemes with a comparison of the post-quantum signatures in Table~\ref{table:pqsystem}.
There are two generic paradigms to obtain the lattice-based signature schemes~\cite{SAC:EFGT16}, one is \textit{hash-and-sign signature} paradigm, 
\eg, Ducas-Lyubashevsky-Prest signature~\cite{AC:DucLyuPre14}, and another is \textit{Fiat-Shamir signature} paradigm, such as 
G\"{u}neysu-Lyubashevsky-P\"{o}ppelmann~\cite{CHES:GunLyuPop12} and BLISS~\cite{C:DDLL13}. Unfortunately for them, these schemes are less efficient and without the unique property.

Notably,  
the current quantum-resistant hash-based signatures, such as SPHINCS\footnote{https://sphincs.cr.yp.to/papers.html}~\cite{EC:BHHLNPS15}, SPHINCS$^{+}$\footnote{https://sphincs.org/resources.html}~\cite{SPHINCS+}, and EPID signature~\cite{CTRSA:BonEskFis19} \etc, cannot satisfy practical performances and cannot provide the public verifiability. Remarkably, Picnic is the first quantum-safe signature with public verifiability from symmetric-key proposed by Chase \etal~\cite{CCS:CDGORR17} using the ``Play-MPC-in-the-Head'' paradigm~\cite{STOC:IKOS07}. But the signature size of Picnic\footnote{https://microsoft.github.io/Picnic/} is near to 40 KBytes. Thus, various optimizations are proposed to shorten the length of keys and signatures while improving the computation performance~\cite{CCS:KatKolWan18}. 
Notably, very recently, an independent and similar result with us was proposed by Buser \etal~\cite{EPRINT:BDEKKL21}, they used the XMSS signature as the building block to design quantum-secure VRF, their methodology is using the unique signature (\eg, XMSS signature) as the beginning point.


\subsection{Revisiting Random Beacons}
The need to generate a large number of high-quality random numbers
is a largely unrecognized requirement of a production blockchain consensus community.
The public randomness beacon enables to issue periodically (\ie, at regular intervals) fresh unpredictable and unbiased random values that was first proposed by Rabin~\cite{JCSS:Rabin83} for contract signing. Afterwards, several applications for cryptographic primitives, such as anonymous browsing~\cite{USENIX:DinMatSyv04}, and blockchain~\cite{Algorand16,DFINITY}, \etc, have been invented. 
The original idea of Rabin cannot work without supporting a trusted third party. Thus, a distributed random beacon that could guarantee output delivery and uniformly distribute the randomness for participants is requested. Recently, some elegant distributed random beacon schemes are proposed for blockchain consensus protocols, such as Ouroboros~\cite{C:KRDO17} and HydRand~\cite{SP:SJSW20} supported by public verifiable secret sharing (or PVSS), Algorand~\cite{Algorand16}, Ouroboros Praos~\cite{EC:DGKR18}, and decentralized random beacon \cite{EPRINT:GLOW20} are armed with $\VRF$s. However, the aforementioned solutions cannot remain secure in the quantum computing era. Designing a post-quantum random beacon is increasingly attracted to researchers' interest. 

\section{Preliminaries and Building Blocks}\label{sec:Preliminaries}



In this section, we review and list definitions and notations that will be used in the following principal contents.
We begin by denoting $\secp$ as the security parameter, then we review the standard definition of the pseudo-randomness functions $\hashH(\cdot)$, $\hashH_1(\cdot)$, $\hashH_2(\cdot)$ and $\hashH_3(\cdot)$  in $\poly(\secp)$ time.
Apart from the well known pseudo-randomness property, the key one-way states that it is hard to find a $\signkey$ such that the function $\hashF_{\signkey}$ maps a given input $x$ to a given output $y$.
$\hashF_{\signkey}$ is instantiated by SHA-$256$ that provides $128$ bits of pre-image resistance against quantum adversaries. 


\par\noindent\textbf{(Non-interactive) Zero-Knowledge (NIZK)}.
For every $\secp$, $\calR$ is denoted as an efficiently computable binary relation, for any pair $(x,w) \in \calR$, we denote $\calL_\calR$ as the language consisting of statements in $\R$, \ie~$\calL_\calR=\{x|\exists~w
\,\, {\rm s.t. } \,\, (x,w)\in \R \}$, where $x$ is a \emph{statement} and $w$ is the associated \emph{witness}.
In other words, there is a polynomial time algorithm running in $\poly(\secp)$ that decides whether $(\stmt,\wits)\in \calR_{\secp}$.
A Sigma protocol for $\calR_{\secp}$ consists of two polynomial time algorithms, prover $P$ and verifier $V$, and proceeds as follows. 
\begin{itemize}
  \item $(\aaa, \stt)\gets\Com(1^{\secp}, \sk:=(\stmt, \wits))$ is executed by $P$, and he keeps the corresponding  state $\stt$ privately and sends the commitment $\aaa$ to the verifier $V$.
  \item a challenge $\ccc$ sampled at random in $\set{0,1}^{\secp}$ is generated by $V$, given $\stmt$ and $\aaa$. 

  \item $\zzz\gets \Prove(1^{\secp}, \sk:=(\stmt, \wits), \stt, \ccc)$ is executed by $P$ given $\ccc$, and outputs a response $\zzz$.

  \item $V$ executes $\Verify(1^{\secp}, \pk:=\stmt, \aaa, \ccc, \zzz)$ checks  whether $(\aaa, \ccc, \zzz)$ is valid. If it is, $\Verify(1^{\secp}, \pk:=\stmt, \aaa, \ccc, \zzz)$ returns 1.
\end{itemize}
When it is clear in the context, we omit $1^{\secp}$ for convenience. In this setting, $\pk$ is denoted as $\stmt$ and $\sk$ is denoted as $(\stmt, \wits)$.
Additionally, a NIZK of knowledge can be transformed from any public-coin zero-knowledge proof (\eg, Sigma-protocol) by using the Fiat-Shamir transform.

\noindent{\textbf{Verifiable Random Function.}}
A \VRF~\cite{FOCS:MicRabVad99} consists of algorithms $(\VrfGen, \VrfEval, \VrfVrfy)$ and satisfies the following syntaxes. 
\begin{itemize}
\item $\VrfGen(1^{\secp})$ outputs a pair of keys $(\vpk, \vsk)$;
\item $\VrfEval(\vsk, \vrfmsg)$ outputs a pair $(\vrfout, \vrfpf)$, where $\vrfout$ is the output value from evaluation function $\Eval(\vsk , \vrfmsg)$, and $\vrfpf$ is the proof of correctness given $\vsk$, \ie, $\vrfpf\gets \Prove(\vsk, \sigma)$. 
\item  $\VrfVrfy(\vpk,\vrfout,\vrfmsg,\vrfpf)$ verifies that $\vrfout=\Eval(\vsk, \vrfmsg)$ using $\vrfpf$, then return 1 if $\vrfout$ is valid and 0 otherwise.
\end{itemize}
Additionally, we require the property of \textit{uniqueness}, \textit{provability} and \textit{pseudorandomness}. \textit{Uniqueness} implies that no values $(\vpk, \vrfmsg, \vrfout_1, \vrfout_2, \vrfpf_1, \vrfpf_2)$ can satisfy the equation $\VrfVrfy(\vpk, \vrfout_1, \vrfmsg, \vrfpf_1)=\Vrfy(\vpk,\vrfout_2,  \linebreak \vrfmsg, \vrfpf_2)$ when $\vrfout_1\neq \vrfout_2$. In other words, for every $\vrfmsg$, only a unique value $\vrfout=\Eval(\vsk, \vrfmsg)$ enables to pass the verification. 
\textit{Provability} implies that  $\VrfVrfy(\vpk, \vrfmsg, \vrfout, \vrfpf)=1$ if $(\vrfout, \vrfpf)=\VrfEval(\vsk,\vrfmsg)$ and $\vrfpf$ is computable given $\vsk$. 
\textit{Pseudorandomness} means that for any probabilistic polynomial-time (PPT) adversary, it is hard to distinguish the function values from
real random ones, which can be guaranteed by hashing function $\Eval(\cdot)$.
Formally, for any PPT adversary $\attacker{A}=(\attacker{A}_1, \attacker{A}_2)$ who did not call its oracle on message $\vrfmsg$, the following probability is at most $1/2+\negl(\secp)$ for some negligible function $\negl(\cdot)$ in the security parameter $\secp$,
\begin{equation*}
\Pr\left[b=b'\middle|
\begin{array}{ll}
 (\vpk,\vsk)\gets \VrfGen(1^{\secp}); &
 (\vrfmsg, \st)\gets \attacker{A}_1^{\VrfEval(\cdot)}(\vpk);\\
 b\gets\set{0,1};  &
\vrfout_0\gets \Eval(\vsk,\vrfmsg);\\
\vrfout_1\gets \set{0,1}^{b(\secp)}; &
b'\gets \attacker{A}_2^{\VrfEval(\cdot)}(\vrfout_b,\st).
\end{array}
\right].
\end{equation*}

\begin{remark}
We remark that, in general, the \textit{post-quantum \VRF}  could be achieved if the cryptographic building blocks are quantum security. We omit to introduce the quantum adversary with the advantages to capture the quantum-secure definition. 

\end{remark}

\par\noindent\textbf{Distributed-\VRF}. Below, the definition is adopted from~\cite{EPRINT:GLOW20}.
In setup phase, $n$ servers $S_1, S_2, \cdots, S_n$ communicate via pairwise private and authenticated channels. They have access to an append-only public board where every server can post messages, and these posts cannot be repudiated by their senders.
A setup interaction is then run between $n$ servers to build a global public key $\pk$, individual servers' public verification keys $\vpk_1, \vpk_2, \cdots, \vpk_n$, and individual servers' secret key $\vsk_1, \vsk_2, \cdots, \vsk_n$. The servers' secret and verification keys $(\vsk_i, \vpk_i)$ for $i=1,2,\cdots,n$ will later enable any subset of $t+1$ servers to non-interactively 
compute the verifiable random value $\Eval(\vsk, \vrfmsg)$ on a plaintext $\vrfmsg\in \calD$. On the contrary, any set of at most $t$ servers cannot learn any information on  $\Eval(\vsk, \vrfmsg)$ for any $\vrfmsg$ not previously computed.
A $(t, n)$ (non-interactive) distributed-\VRF consists of the following algorithms $(\DistKG, \PEval, \Comb, \Vrfy)$: 
\begin{itemize}
\item $(\vpk, \set{\vvk_i}_{i\in [n]}, \set{\vsk_i}_{i\in [n]})\gets\DistKG(1^{\secp}, t, n)$ is a fully distributed key generation algorithm that takes as input a security parameter $1^{\secp}$, the number of participating servers $n$, and the threshold parameter $t$; it outputs a set of qualified servers $\textrm{QUAL}$, a global public key $\vpk$, a list $\set{\vsk_1, \cdots, \vsk_n}$ of server's secret keys, a list $\set{\vvk_1, \cdots, \vvk_n}$ of servers' verification keys.

\item $(\vrfout_i, \vrfpf_i)\gets\PEval(\vsk_i, \vvk_i, \vrfmsg)$ is a partial evaluation algorithm that takes as input server $S_i\in \textrm{QUAL}$, secret key $\vsk_i$, and verification key $\vvk_i$, a plaintext $\vrfmsg$, and outputs either a triple $\vrfpf_i=(i, \vrfout_i, \strresp_i, \vrfchlg_i)$, where $\vrfout_i$ is the $i$-th evaluation share and $\strresp_i$ is a non-interactive proof of correct partial evaluation.

\item $(\vrfout, \vrfpf) \gets \Comb(\vpk, \vvk, \vrfmsg, \bigepsilon)$ is a combination algorithm that takes as input the global public key $\vpk$, the verification key $\vvk$, a message $\vrfmsg$, and a set $\bigepsilon=\set{\vrfpf_{i_1}, \cdots, \vrfpf_{i_{\abs{\bigepsilon}}}}$ of partial function evaluations origination from $\abs{\bigepsilon}\geq t+1$ different servers, and outputs either a pair $(\vrfout, \vrfpf)$ of pseudo-random function value $\vrfout$ and correctness proof $\vrfpf$, or $\bot$.

\item $\set{0,1}\gets\Vrfy(\vpk, \vvk, \vrfmsg, \vrfout, \vrfpf)$ is a verification algorithm that takes as input the public key $\vpk$, a set of verification key $\vvk$, a plaintext $\vrfmsg$, and a proof $\vrfpf$, then outputs $1$ or $0$ (\ie, accept or reject).
\end{itemize}

Additionally, the distributed \VRF satisfies the following properties. 
\begin{enumerate*}
\item \textit{Consistency}, meaning that no matter which collection of correctly offered shares is used to compute the function on a plaintext $\vrfmsg$ the same random value $\vrfout=\hashF(\vsk, \vrfmsg)$ is obtained.
\item \textit{Domain-range correctness}, meaning that every computed value $\vrfout$ belongs to the range domain $\calR$.
\item \textit{Probability} means that the uniquely recovered value $\vrfout=\hashF(\vsk, \vrfmsg)$ passes the verification test.

\item \textit{Uniqueness} means that for every plaintext $\vrfmsg$ a unique $\vrfout=\hashF(\vsk, \vrfmsg)$  passes the verification.
\end{enumerate*}

\par\noindent\textbf{Linear Decomposition of a Circuit}. 
ZKP schemes atop MPC-in-the-head paradigm~\cite{STOC:IKOS07} uses an explicit linear $(2,3)$-decomposition of a circuit $\phi: \R^{m} \to \R^{\ell}$ over an arbitrary finite ring $\R$. Here $\phi$ can be expressed by an $n$-gate arithmetic circuit over the ring, 
 and supports multiplication by constant, binary addition and binary multiplication gates.
Below, the linear $(2,3)$-decomposition of a circuit $\phi$ is defined for the statement $\stmt=\phi(\wits)$ with a witness $\wits$.
\begin{itemize}
\item $(\wits_1, \wits_2, \wits_3)\gets \Share(\wits, \tape_1, \tape_2, \tape_3)$. Picks random tapes $\tape_1, \tape_2, \tape_3 \in \R^m$. Output $\wits_1=\hashG_1(0\cdots\abs{\wits-1})$, $\wits_2=\hashG_2(0\cdots\abs{\wits-1})$, and $ \wits_3 = \wits - \wits_1 -\wits_2$ for a pseudorandom generator $\hashG_i$ seeded with $\tape_i$,  where $\hashG_i(0\cdots\abs{\wits-1})$ outputs the first $\abs{\wits}$-bit.

\item $ \zkbviewxx{i+1}_j \gets \Updatec(\zkbviewi_j , \zkbviewi_{j+1}, \tapei_j, \tapei_{j+1})$, $j\in [1, n]$ and $i\in [1,t]$. Takes as input the $\zkbviewi_j$ and $\tapei_j$ of the participant $P_j$ as well as $\zkbviewi_{j+1}$ and $\tapei_{j+1}$ of $P_{j+1}$.
Computes the participant $P_j$'s view of the output wire of gate $\gate_i$ and appends it to the view. $\wire_\delta$ is denoted as $\delta$-th wire, and $\wire^{(j)}_\delta$ is referred to the value of $\wire_k$ in the view of $P_j$. It is notable that 
the $\Updatec$ operation depends on the type of gate $\gate_j$ and supports the operation of addition by constant, 
 multiplication by constant, binary addition and binary multiplication gates.
\begin{itemize}
\item \textit{Addition by constant} ($\wire_b = \wire_a + \delta$). Outputs $\wire^{(j)}_b = \wire^{(j)}_a + \delta$ if $j=1$. Otherwise, outputs $\wire^{(j)}_b = \wire^{(j)}_a$.
\item \textit{Multiplication by constant} ($\wire_b = \delta\cdot\wire_a$). Outputs  $\wire^{(j)}_b = \delta\cdot \wire^{(j)}_a$.
\item \textit{Binary addition}  ($\wire_c = \wire_a +  \wire_b$). Outputs $\wire^{(j)}_c = \wire^{(j)}_a + \wire^{(j)}_b$.
\item \textit{Binary multiplication} ($\wire_c = \wire_a \cdot \wire_b$). Outputs $\wire^{(j)}_c = \wire^{(j)}_a \cdot \wire^{(j)}_b + \wire^{(j+1)}_a \cdot \wire^{(j)}_b + \wire^{(j)}_a \cdot \wire^{(j+1)}_b + \PRG_j(c) - \PRG_{j+1}(c)$, where $\PRG_j(c)$ is the $c$-th output seeded with $\tape_j$.
\end{itemize}
\item $\zkbshare_j \gets \Output(\zkbviewxx{n}_j)$. Selects the $\ell$ output wires of the circuit as stored in $\zkbviewxx{n}_j$.

\item $\stmt \gets \Reconst(\stmt_1,\stmt_2,\stmt_3)= \zkbshare_1 + \zkbshare_2 + \zkbshare_3$. Reconstructs 
$\stmt$.

\end{itemize}


\noindent\textbf{Non-Interactive ZKBoo atop MPC-in-the-head Paradigm}.
Ishai-Kushilevitz-Ostrovsky-Sahai (IKOS)~\cite{STOC:IKOS07} proved a surprising result in 2007 that even semi-honest multiparty computation (MPC) is sufficient to obtain ZKPs.  Additionally, IKOS paradigm~\cite{STOC:IKOS07} enables to provide a transformation to obtain a zero-knowledge protocol from symmetric primitives with low communication complexity. As we know, 
$\ZKBoo$ and $\ZKBpp$ build on the MPC-in-the-head paradigm of Ishai \etal~\cite{STOC:IKOS07},
%
%
%
%
And we conclude their main ideas as follows.  For the public $\phi$ and the word $\word\in \calL_{\phi}$, the prover is with the witness $\wits$ such that $\word = \phi(\wits)$. In addition, $\Com(\cdot)$ is denoted as the commitment, $\hashH(\cdot)$ is the public hash function for the prover and verifier, and the integer $t$ is the number of parallel iterations.


\begin{itemize}
\item $\strresp \gets\Prove_{\hashH}(\strstmt:=\word, \strwits:= \wits)$. The prover simulates an MPC protocol ``in their head'', commits to the state and transcripts of all players. More concretely,
\begin{enumerate}
\item For $i=1$ to $t$, set $r_i$ as the iteration label, and for $j=1$ to $3$, to obtain an output $\zkbviewi_j$ and share $\zkbsharei_j$,the player $P_j$ picks random tapes $\tapei_1, \tapei_2, \tapei_3$ and de-composites the witness $\wits$ as follows: 
\begin{enumerate}
\item 
shares $(\zkbwitsi_1, \zkbwitsi_2, \zkbwitsi_3)\gets \Share(\wits, \tapei_1, \tapei_2, \tapei_3)$;
\item computes the $i$-th view of participant $j$, \ie, $\zkbviewi_j \gets \Updatec(\Updatec(\cdots\Updatec(\zkbwitsi_j, \zkbwitsi_{j+1}, \tapei_j, \tapei_{j+1})\linebreak \cdots)\cdots)$, and outputs partial result $\zkbsharei_j \gets \Output(\zkbviewi_j)$;
\item finally commits $(\ccti_j, \copi_j)\gets \Com(\tapei_j, \zkbviewi_j)$ and sets $\mstmti = (\zkbsharei_1, \zkbsharei_2, \zkbsharei_3, \ccti_1, \ccti_2, \ccti_3)$.
\end{enumerate}


\item $P_j$ computes the challenge $\chlgc \gets \hashH(\mstmtxx{1}, \mstmtxx{2}, \cdots, \mstmtxx{t})$, where $\chlgc$ can be interpreted as $\chlgc^{(i)} \in \{1,2,3\}$ for $i=[1, t]$.

\item For $i=1$ to $t$, set $r_i$ as the iteration label, the player $P_j$ creates $z^{(i)} = (\copi_\chlgc, \copi_{\chlgc+1})$. Then the player $P_j$ outputs $\strresp = \left((\mstmtxx{1}, z^{(1)}), (\mstmtxx{2}, z^{(2)}), \cdots, (\mstmtxx{t}, z^{(t)})\right)$.
\end{enumerate}

\item $b \gets \Verify_{\hashH}(\strstmt:=\word, \strresp)$. The verifier ``corrupts'' a random subset of the simulated players by seeing their complete state, then the verifier checks that the computation was done correctly from the perspective of the corrupted players, and if so, the verifier has some assurance that the output is correct and the prover knows $x$. Iterating this for many rounds, then the verifier gets high assurance.

\begin{enumerate}
\item For $i=1$ to $t$, set $r_i$ as the iteration label,  the verifier computes the challenge $\chlgc' \gets \hashH(\mstmtxx{1}, \mstmtxx{2}, \cdots, \mstmtxx{t})$, where $\chlgc'$ can be interpreted as $\chlgc'^{(i)} \in \{1,2,3\}$.
Then the verifier validates if there exists $j\in \set{\chlgc'^{(i)}, \chlgc'^{(i)}+1}$ such that $\Open(\ccti_j, \copi_j) = \bot$, outputs \textit{Reject}. Otherwise, for all $j\in \set{\chlgc'^{(i)}, \chlgc'^{(i)}+1}$, sets $\set{\tapei_j, \zkbviewi_j} \gets \Open(\ccti_j, \copi_j)$.

\item Next, the verifier validates $\Reconst(\zkbsharei_1, \zkbsharei_2, \zkbsharei_3) \overset{?}{=} \word$. If the validation does not pass, then outputs \textit{Reject}. If there exists  $j\in \set{\chlgc'^{(i)}, \chlgc'^{(i)}+1}$, then validates  $\zkbsharei_j \overset{?}{=} \Output(\zkbviewi_j)$. If the validation does not pass, then outputs \textit{Reject}. For each wire value $\wire_j^{(e)} \in \zkbview_e$, if $\wire_j^{(e)} \neq \Updatec(\zkbview^{(j-1)}_e, \zkbview^{(j-1)}_{e+1}, \tape_e, \tape_{e+1})$ output \textit{Reject}.

\item The verifier outputs \textit{Accept}.
\end{enumerate}
\end{itemize}

\section{Post-Quantum Verifiable Random Function}
\label{sec:framework}
An observation is that Algorand does not adopt the Dodis-Yampolskiy \VRF that depends on the costly bilinear pairing and cannot service the large-scale PoS network. Instead, Algorand
recommends to adopt the Goldberg-Naor-Papadopoulos-Reyzin \VRF~\cite{EPRINT:GNPR16} in his realization that bypasses the unique signature as intermediation.
Thus, to prevent quantum computer attacks while servicing the large-scale lottery-based PoS consensus protocol (\eg, Algorand), in this work, we present an instance of quantum secure \VRF from symmetric primitives for random beacon and lottery-based PoS consensus protocol; however, unlike previous construction, we realize it in a quantum resilient way, \ie, the proposed \VRF isn't based on hard problems from number theory.

\par\noindent\textbf{Quantum-secure signature}. Before presenting our quantum-secure \VRF construction, we first present 
an indirect approach using signatures  that have the ``uniqueness'' property.
Inspired by the spirit of hash-based \textit{signature from Fiat-Shamir} for Schnorr signature, it is easy to design a quantum-safe signature scheme by integrating a (weakly complete) quantum-secure identification protocol (or $\Sigma$ protocol), the methodology of the signature from Fiat-Shamir is  summarized as follows.     

\begin{itemize}
  \item $\Gen(1^{\secp})$, $\sk:=\Fishsrt \gets R$, and $\vk:=\Fishpk=\OWF(\sk)$.
  \item $\Sign(\sk,\msg)$, it takes $\sk$ and a message $\msg$ as input, and generates
  \begin{enumerate*}
    \item $(\aaa, \stt)\gets \Com(\sk:=\Fishsrt, \opstr:=\rand)$,

    \item $c\gets \textcolor{black}{\hashH(\aaa\|\msg)}$, where $\|$  denotes concatenation.
    \item and $\zzz\gets\Prove(\sk:=\Fishsrt,\stt, \ccc)$.
\end{enumerate*}
  If $\zzz$ is not valid, it runs another round. It keeps running until $\zzz$ is valid. Finally it returns the signature $\sigma:=(\aaa, \ccc, \zzz)$.

  \item   $\Vrfy(\vk, \msg, \sigma)$, it rejects if there exists $\Vrfy(\vk:=y, \msg, (\aaa,\zzz,\textcolor{black}{ \hashH(\msg \| \aaa)}))=0$.
\end{itemize}

In short, the OWF family $\myset{f_k}_{k\in K_{\secp}}$ is used for key generation in both signature schemes, the public key is an image $y=f(x)$ of a one-way function $f$ and secret key $x$, and a signature is an \NIZK proof $\zkproof$ of $x$, that incorporates a message to be signed.

\subsection{Provably Secure VRF from Symmetric Primitives}

Notably, there are two issues that may prevent some kinds of signature schemes with uniqueness from being used straightforwardly as a \VRF.
Firstly, the signature may not be unique by given the message $\msgx$ and the public key $\pk$. Secondly, the signature $\sigma_{\msgx}$ is unpredictable but not pseudorandom (\eg, signatures could contain some bias and be distinguishable from a random distribution.)
In addition, \VRF derived from unique signatures presents strong unbiasibility properties due to the uniqueness, even in the presence of active adversaries, of the corresponding pseudorandom value.
Thus, it is not enough to obtain a \VRF with uniqueness and pseudorandomness from the quantum-secure unique signature in a fairly straightforward way. Indeed, according to our investigation~\cite{NSEC5}\footnote{\url{https://www.cs.bu.edu/~goldbe/projects/vrf}}, there is a direct approach to construct $\VRF$, which can bypass the unique signature as intermediation.

As mentioned earlier, any \VRF has three components, generator, evaluator, and verifier. The evaluator executes two functions, the evaluation and the prover. The paradigm of post-quantum \VRF is achieved by leveraging concepts together with symmetric cryptography and quantum-safe ZKP systems to create a signature. In a nutshell,
this paper introduces a new post-quantum \VRF by using the ideas of~\cite{EPRINT:GNPR16} for construction of the evaluation function and ideas of~\cite{EC:KilLyuSch18} for construction of the prover based on a quantum-safe ZKP for the \textit{uniqueness} property.
 An important difference between this proposed \VRF construction  and~\cite{EPRINT:GNPR16,EC:KilLyuSch18} is the fact that in the proposed \VRF both the evaluator, and verifier should have access to the same secret key. Thus, the use of a priori post-quantum symmetric key protocol is also required.  

To make it understand easily, our \VRF evaluation function uses deterministic hash functions to hash the message and appends a quantum-safe proof of knowledge for the witness secret key, where the prover can convince the verifier that she knows a secret key without disclosing the secret key itself.
There is a public key $\vpk$ associated with a secret seed $\vsk$ and a pesudorandom hash function $\hashF(\cdot)$ with a random key $\rkey\gets \calK$ , then it satisfies
\begin{enumerate*}
  \item $(\vrfout, \vrfpf)=\VrfEval(\vsk,\msgm)$ is efficiently computable given the corresponding $\vsk$, where 
  $\vrfout$ is calculated by using hash functions $\hashH_2(\vsk, \hashH_1(\msgm))$.
  \item A proof $\vrfpf$ is computable given the $\vsk$, and a natural approach is the quantum-safe ZKP by invoking $\Prove(\vrfout, \vsk)$.
  \item No adversary can distinguish $\vrfout$ by computing $\hashH_2(\vsk, \hashH_1(\msgm))$  from a random value
without explicitly querying for $\msgm$.
  \end{enumerate*}

\noindent\textbf{Equality of \ZKBoo}. In addition, to guarantee the uniqueness without disclosing the secret key, a equality of quantum-secure $\ZKP$ system is used to prove that the public key and the output of \VRF have same pre-image secret key here.  Below, we present the \textit{equality} of $\ZKBoo$ system as a warm-up, which is an independent contribution. Here, regarding the relation $\calL=\myset{(\Ourpk, \rkey, \Oursign, \hashh; \Oursk):\Ourpk=\hashF({\rkey}, \Oursk)$, $\Oursign = \hashH_2(\Oursk, \hashh)}$, the equality of \ZKBoo proceeds relations $\calL_{\hashF_1}=\myset{(\Ourpk, \rkey; \Oursk):\Ourpk=\hashF({\rkey},\Oursk)}$ and $\calL_{\hashH_2}=\myset{(\Oursign, \hashh; \Oursk): \Oursign = \hashH_2(\Oursk, \hashh)}$ synchronously. In particular, the prover proceeds as depicted in Fig.\ref{fig:eq-zkboo} and the verifier proceeds as depicted in Fig.~\ref{fig:eq-v-zkboo}.

\mybox{Equality Proof Algorithm of \ZKBoo.}{white!40}{white!10}{

\begin{itemize}
\item $\strresp \gets \ZKP.\Prove_{\hashH}(\strstmt, \strwits)$ 
for the iteration label $i=1$ to $t$ and the dummy participant label $j=1$ to $3$.
\begin{enumerate}
        \item compute $(\Oursk^{(i)}_{\Ourpk,1},\Oursk^{(i)}_{\Ourpk,1}, \Oursk^{(i)}_{\Ourpk,3})\gets \Share(\Oursk, \Ourpk.\tapei_1, \Ourpk.\tapei_2, \Ourpk.\tapei_3)$  and  $(\Oursk^{(i)}_{\Oursign,1},\Oursk^{(i)}_{\Oursign,1}, \Oursk^{(i)}_{\Oursign,3})\gets \Share(\Oursk, \Oursign.\tapei_1, \Oursign.\tapei_2, \Oursign.\tapei_3)$
         for random tapes $\Ourpk.\tapei_{j}$ and $\Oursign.\tapei_{j}$.
        
        \item compute
        $\Ourpk.\zkbviewi_j \gets \UpdateF(\UpdateF(\cdots\UpdateF(\Oursk^{(i)}_j, \Oursk^{(i)}_{j+1}, \Ourpk.\tapei_j, \Ourpk.\tapei_{j+1})\cdots)\cdots)$ and compute 
        $\Oursign.\zkbviewi_j \gets \UpdateH(\UpdateH(\cdots\UpdateH(\Oursk^{(i)}_j, \Oursk^{(i)}_{j+1}, \Oursign.\tapei_j, \Oursign.\tapei_{j+1})\cdots)\cdots)$ (see \ZKBoo for details); 
        \item obtain $\Ourpk.\strstmt^{(i)}_j \gets \Output(\Ourpk.\zkbviewi_j)$ and $\Oursign.\strstmt^{(i)}_j \gets \Output(\Oursign.\zkbviewi_j)$;
        
        \item commit $(\Ourpk.\ccti_j, \Ourpk.\copi_j)\gets \Com(\Ourpk.\tapei_j, \Ourpk.\zkbviewi_j)$ and $(\Oursign.\ccti_j, \Oursign.\copi_j)\gets \Com(\Oursign.\tapei_j, \Oursign.\zkbviewi_j)$;
        
        \item create $\Ourpk.\mstmtxx{i} = (\Ourpk.\strstmt^{(i)}_1, \Ourpk.\strstmt^{(i)}_2, \Ourpk.\strstmt^{(i)}_3, \Ourpk.\ccti_1, \Ourpk.\ccti_2, \Ourpk.\ccti_3)$ and  $\Oursign.\mstmtxx{i} = (\Oursign.\strstmt^{(i)}_1, \Oursign.\strstmt^{(i)}_2, \Oursign.\strstmt^{(i)}_3, \Oursign.\ccti_1, \Oursign.\ccti_2, \Oursign.\ccti_3)$;  
           
        \item compute the challenge internally $\strchlg\in\myset{1,2,3}$, \ie, 
        $\chlgc \gets \hashH(\Ourpk.\mstmtxx{1}, \Ourpk.\mstmtxx{2}, \cdots, \Ourpk.\mstmtxx{t}\| \Oursign.\mstmtxx{1}, \Oursign.\mstmtxx{2}, \cdots, \Oursign.\mstmtxx{t})$;

        \item create $\Ourpk.z^{(i)} = (\Ourpk.\copi_\chlgc, \Ourpk.\copi_{\chlgc+1})$ and $\Oursign.z^{(i)} = (\Oursign.\copi_\chlgc, \Oursign.\copi_{\chlgc+1})$;

        \item output $\Ourpk.\strresp = \left((\Ourpk.\mstmtxx{1}, \Ourpk.z^{(1)}), (\Ourpk.\mstmtxx{2}, \Ourpk.z^{(2)}), \cdots, (\Ourpk.\mstmtxx{t}, \Ourpk.z^{(t)})\right)$ and $\Oursign.\strresp = \left((\Oursign.\mstmtxx{1}, \Oursign.z^{(1)}), (\Oursign.\mstmtxx{2}, \Oursign.z^{(2)}), \cdots, (\Oursign.\mstmtxx{t}, \Oursign.z^{(t)})\right)$.
\end{enumerate}
\end{itemize}

}{Equality Proof Algorithm of \ZKBoo.~\label{fig:eq-zkboo}}

\mybox{Equality Verification Algorithm of \ZKBoo.}{white!40}{white!10}{
\begin{itemize}
\item 
$b \gets \Verify_{\hashH}(\strstmt:=\word, \strresp)$ for the iteration label $i=1$ to $t$ and the dummy participant label $j=1$ to $3$.
\begin{enumerate}     
\item parse $\Ourpk.\strresp$ into the sequence of $\Ourpk.\mstmtxx{i}$ and $\Ourpk.z^{(i)}$, and parse $\Oursign.\strresp$ into $\Oursign.\mstmtxx{i}$ and $\Oursign.z^{(i)}$ sequences;
 \item compute the challenge $\chlgc' \gets \hashH(\Ourpk.\mstmtxx{1}, \Ourpk.\mstmtxx{2}, \cdots, \Ourpk.\mstmtxx{t}\| \Oursign.\mstmtxx{1}, \Oursign.\mstmtxx{2}, \cdots, \Oursign.\mstmtxx{t})$;
 \item validate $\forall~j\in \myset{\chlgc'^{(i)}, \chlgc'^{(i)}+1}$, then set $\myset{\Ourpk.\tapei_j, \Ourpk.\zkbviewi_j} \gets \Open(\Ourpk.\ccti_j, \Ourpk.\copi_j)$ and 
 $\myset{\Oursign.\tapei_j, \Oursign.\zkbviewi_j} \gets \Open(\Oursign.\ccti_j, \Oursign.\copi_j)$;
 
 \item validate $\Reconst(\Ourpk.\strstmt^{(i)}_1, \Ourpk.\strstmt^{(i)}_2, \Ourpk.\strstmt^{(i)}_3) \overset{?}{=} \Ourpk.\strstmt$ and $\Reconst(\Oursign.\strstmt^{(i)}_1, \Oursign.\strstmt^{(i)}_2, \Oursign.\strstmt^{(i)}_3) \overset{?}{=} \Oursign.\strstmt$;
 
 \item validate $\exists~j\in \myset{\chlgc'^{(i)}, \chlgc'^{(i)}+1}$, $\Ourpk.\strstmt^{(i)}_j \overset{?}{=} \Output(\Ourpk.\zkbviewi_j)$ and $\Oursign.\strstmt^{(i)}_j \overset{?}{=} \Output(\Oursign.\zkbviewi_j)$;

 \item validate $\forall~\Ourpk.\wire_j^{(e)} \in \Ourpk.\zkbview_e$ and $\forall~\Oursign.\wire_j^{(e)} \in \Oursign.\zkbview_e$, then validate
  $\Ourpk.\wire_j^{(e)} \overset{?}{=} \UpdateF(\Ourpk.\zkbview^{(j-1)}_e, \Ourpk.\zkbview^{(j-1)}_{e+1}, \Ourpk.\tape_e, \Ourpk.\tape_{e+1})$ and $\Oursign.\wire_j^{(e)} \overset{?}{=} \UpdateF(\Oursign.\zkbview^{(j-1)}_e, \Oursign.\zkbview^{(j-1)}_{e+1}, \Oursign.\tape_e, \Oursign.\tape_{e+1})$;
 \item output \textit{1} if all validations are passed.
\end{enumerate}  
\end{itemize}

}{Equality Verification Algorithm of \ZKBoo.~\label{fig:eq-v-zkboo}}




\noindent\textbf{Post-Quantum \VRF via \ZKBoo}. Armed with the equality of \ZKBoo, the detailed post-quantum \VRF via \ZKBoo is depicted in Fig.~\ref{fig:our-vrf-gen} for key generation, Fig.~\ref{fig:our-vrf-eval} for evaluation, and Fig.~\ref{fig:our-vrf-vrfy} for verification.
Notably, we only present a post-quantum \VRF via \ZKBoo. Indeed, it is easy to obtain a quantum-secure \VRF via other candidates post-quantum $\ZKP$, such as \ZKBpp and ZKSTARK, but we ignore the details here.

\mybox{$\VrfGen$ of Post-Quantum \VRF via ZKBoo.}{white!40}{white!10}{
\begin{itemize}
\item $(\vsk, \vpk)\gets\VrfGen(1^{\secp})$.

\begin{enumerate}
  \item Choose a secret key $\Oursk \gets \calD$ and a random key $\rkey\gets \calK$ for the pseudo-random function $\hashF$, and return $\vsk \gets (\rkey, \Oursk)$.
  \item Compute the public key $\vpk:=\Ourpk=\hashF({\rkey}, \vsk)=\hashF({\rkey}, \Oursk)$. 
\end{enumerate}

\end{itemize}
    
}{Key Generation of Post-Quantum\VRF via ZKBoo.~\label{fig:our-vrf-gen}}

\mybox{$\VrfEval$ of Post-Quantum \VRF via ZKBoo.}{white!40}{white!10}{
\begin{itemize}
\item $(\vrfout, \vrfpf)\gets\VrfEval(\vsk:=(\rkey,\Oursk), \msg)$.
\begin{enumerate}
\item Compute $\hashh=\hashH_1(\msgm)$, $\Oursign = \hashH_2(\Oursk, h)$, and $\beta=\hashH_3(\Oursign)$.

  
  \item Regarding $\Ourpk=\hashF({\rkey}, \Oursk)\in \calL_{\hashF}$ for a circuit $\hashF$, and  $\Oursign = \hashH_2(\Oursk, h)  \in \calL_{\hashH_2}$ for a circuit $\hashH_2$, invoke the quality of post-quantum $\ZKP$ system for the relation $\calL=\myset{(\Ourpk, \rkey, \Oursign, \hashh; \Oursk):\Ourpk=\hashF({\rkey}, \Oursk)$, $\Oursign = \hashH_2(\Oursk, \hashh)}$ by inputting the statement $\strstmt:=(\Ourpk, \rkey, \Oursign, \hashh)$ and the witness $\strwits:=\vsk = \Oursk$, then obtain the proof of knowledge of secret key to prove the output $(\Ourpk, \Oursign)$ is the correct hash output. In particular, invoke 
       \begin{eqnarray*}
       \strresp &\gets & \ZKP.\Prove_{\hashH}(\strstmt, \strwits)
        \end{eqnarray*}
as depicted in Fig.\ref{fig:eq-zkboo} to proceed relations $\calL_{\hashF}=\myset{(\Ourpk, \rkey; \Oursk):\Ourpk=\hashF({\rkey},\Oursk)}$ and $\calL_{\hashH_2}=\myset{(\Oursign, \hashh; \Oursk): \Oursign = \hashH_2(\Oursk, \hashh)}$ respectively, and return a computation integrity proof $\vrfpf=(\strchlg:=\chlgc, \strresp:=(\Ourpk.\strresp, \Oursign.\strresp))$, where internally the challenge $\chlgc$ is involved. 
   

  \item Last return the output of verifiable random function $\vrfout=(\beta, \strstmt)$ and the proof of correctness $\vrfpf$. 
  
\end{enumerate}
\end{itemize}

}{Evaluation of Post-Quantum \VRF via ZKBoo.~\label{fig:our-vrf-eval}}


%
\mybox{$\VrfVrfy$ of Post-Quantum \VRF via ZKBoo.}{white!40}{white!10}{
\begin{itemize}
\item $\VrfVrfy((\vpk, \vrfout), \vrfmsg, \vrfpf)$.
\begin{enumerate}
  \item verify $\hashh \overset{?}{=}\hashH_1(\msgm)$;
  \item invoke the equality verification of \ZKBoo to validate 
  \begin{equation*}
  \ZKP.\Verify_{\hashH}(\strstmt, \vrfpf)
  \end{equation*}
  as depicted in Fig.~\ref{fig:eq-v-zkboo} for relations $\calL_{\hashF}=\myset{(\Ourpk, \rkey; \Oursk):\Ourpk=\hashF({\rkey},\Oursk)}$ and $\calL_{\hashH_2}=\myset{(\Oursign, \hashh; \Oursk): \Oursign = \hashH_2(\Oursk, \hashh)}$ respectively;

  \item  Return $1$ if $\ZKP.\Verify_{\hashH}(\strstmt, \vrfpf)=1$ holds, 
  and 0 otherwise.
    
\end{enumerate}
\end{itemize}
}{Verification of Post-Quantum \VRF via ZKBoo.~\label{fig:our-vrf-vrfy}}

%


\subsection{Post-Quantum VRF Security Analysis}

Importantly, the proposed \VRF isn't based on hard problems from number theory, and
the hard problems the proposed \VRF relies on for security relate only to symmetric cryptographic primitives that are thought to be secure against quantum attacks, and quantum security can be guaranteed easily.
Below, we analyze uniqueness and pseudorandomness sketchily.  

\par\noindent\textbf{Uniqueness Analysis}.
%
%
%
%
%
%
%
%
The uniqueness property requires that there should be only one provable VRF output $\vrfout$ for every input $\vrfmsg$. In particular, for every $\big(\vpk, \vrfmsg, (\vrfout', \vrfpf'),  (\vrfout^*, \vrfpf^*)\big)$ such that $\vrfout' \neq \vrfout^*$, the following 
$\VrfVrfy(\vpk, \vrfmsg, (\vrfout', \vrfpf'))=\VrfVrfy(\vpk, \vrfmsg, (\vrfout^*, \vrfpf^*))$.
We prove the property of uniqueness using a contradiction. 
If there is an adversary that violates computational uniqueness, the adversary can come up with a message $\vrfmsg' (\neq \vrfmsg)$ given $\vsk$. 
The generated statement $\strstmt'$ contains $(\vpk:=\Ourpk, \vrfout':=\Oursign')$ for $\hashh':=\hashH_1(\msgm')$, and proof $\vrfpf'$ contains $\vrfpf'=( \strchlg':=\chlgc', \strresp':=\resp')$ such that an incorrect \VRF output value $\vrfout':=\beta'=\hashH_3(\Oursign')$ is computed for $\Oursign'=\hashH_2(\vsk, \hashh'=\hashH_1(\vrfmsg'))$ and a different message $\vrfmsg'$. 
Note that the correctness output of \VRF $\vrfout$ via $\VrfEval$ is computed as $\vrfout:=\beta=\hashH_3(\Oursign)$ for the corresponding $\Oursign = \hashH_2(\vsk, \hashh:=\hashH_1(\vrfmsg))$. Since $\vrfout'\neq \vrfout$ (\ie, $\beta' \neq \beta$), we have $\Oursign' \neq \Oursign$, where $\Oursign':= \hashH_2(\vsk, \hashh':=\hashH_1(\vrfmsg'))$ and $\Oursign := \hashH_2(\vsk, \hashh:=\hashH_1(\vrfmsg))$ for the same $\vsk$.
Now, $\vrfpf'=(\strchlg', \strresp')$ 
for a challenge $\chlgc'$ and a response $\resp'$ ensures that $\VrfVrfy((\vpk, \vrfout'), \vrfmsg', \vrfpf')=1$ supported by the $\ZKP.\Verify_{\hashH}(\strstmt:=(\Ourpk, {\Oursign'}, \beta'), \vrfpf'=(\chlgc', \resp'))=1$ of the equality of $\ZKBoo$ system. 

$\VrfVrfy(\cdot)$ ensures to extract the exact one challenge 
$$\chlgc' \gets \hashH(\Ourpk.\mstmtxx{1}, \cdots, \Ourpk.\mstmtxx{t}\| \Oursign.\mstmtxxp{1}, \cdots, \Oursign.\mstmtxxp{t})$$
where $\Ourpk.\mstmtxx{i}= (\Ourpk.\strstmt^{(i)}_1, \Ourpk.\strstmt^{(i)}_2, \Ourpk.\strstmt^{(i)}_3, \Ourpk.\ccti_1, \Ourpk.\ccti_2, \Ourpk.\ccti_3)$ and  $\Oursign.\mstmtxxp{i}= (\Oursign.\strstmt'^{(i)}_1, \Oursign.\strstmt'^{(i)}_2, \Oursign.\strstmt'^{(i)}_3,  \Oursign.\cctip_1, \Oursign.\cctip_2, \Oursign.\cctip_3)$. %
Below, we use the extract $\chlgc'$ to perform validations as follows.  
 \begin{enumerate}
 \item Firstly, validate if $\forall~j\in \myset{\chlgc'^{(i)}, \chlgc'^{(i)}+1}$ such that $\myset{\Ourpk.\tapei_j, \Ourpk.\zkbviewi_j} \gets \Open(\Ourpk.\ccti_j, \Ourpk.\copi_j)$ and 
 $\myset{\Oursign.\tapei_j, \Oursign.\zkbviewi_j}  \gets \Open(\Oursign.\ccti_j, \Oursign.\copi_j)$, then continue; Otherwise, if $\exists~j\in \myset{\chlgc'^{(i)}, \chlgc'^{(i)}+1}$ such that $\bot \gets \Open(\Ourpk.\ccti_j, \Ourpk.\copi_j)$ and 
 $\bot \gets \Open(\Oursign.\ccti_j, \Oursign.\copi_j)$, then abort the validation.
 
 \item Then validate $\Reconst(\Ourpk.\strstmt^{(i)}_1, \Ourpk.\strstmt^{(i)}_2, \Ourpk.\strstmt^{(i)}_3) \overset{?}{=} \Ourpk.\strstmt$ and $\Reconst(\Oursign.\strstmt^{(i)}_1, \Oursign.\strstmt^{(i)}_2, \Oursign.\strstmt^{(i)}_3) \overset{?}{=} \Oursign.\strstmt$, if pass the validation, then continue the following validation; otherwise, abort it.
 
 \item Next, validate if $\exists~j\in \myset{\chlgc'^{(i)}, \chlgc'^{(i)}+1}$ such that $\Ourpk.\strstmt^{(i)}_j \overset{?}{=} \Output(\Ourpk.\zkbviewi_j)$ and $\Oursign.\strstmt^{(i)}_j \overset{?}{=} \Output(\Oursign.\zkbviewi_j)$, then continue; otherwise, then abort.

 \item Finally, validate if $\forall~\Ourpk.\wire_j^{(e)} \in \Ourpk.\zkbview_e$ and $\forall~\Oursign.\wire_j^{(e)} \in \Oursign.\zkbview_e$, then validate
  $\Ourpk.\wire_j^{(e)} \overset{?}{=} \UpdateF(\Ourpk.\zkbview^{(j-1)}_e, \Ourpk.\zkbview^{(j-1)}_{e+1},  \Ourpk.\tape_e, \Ourpk.\tape_{e+1})$ and $\Oursign.\wire_j^{(e)} \overset{?}{=} \UpdateF(\Oursign.\zkbview^{(j-1)}_e, \Oursign.\zkbview^{(j-1)}_{e+1}, \Oursign.\tape_e, \Oursign.\tape_{e+1})$;
 \item output \textit{1} if all validations are passed.
\end{enumerate}  
Notably, $\Oursign.\mstmtxxp{i}$ is not identical to $\Oursign.\mstmtxx{i}$ because of $\Oursign':=\hashH_2(\vsk, \hashh'=\hashH_1(\vrfmsg'))  \neq \Oursign:=\hashH_2(\vsk, \hashh=\hashH_1(\vrfmsg))$. In addition, $\hashH_2$ is a random oracle, its output is random, and the probability that it equals the unique value determined by its inputs according that the right side of equation~(\ref{eq:extract-pq-chlg}) is negligible. 
\begin{eqnarray}\label{eq:extract-pq-chlg}
\nonumber
\chlgc &= &\hashH(\Ourpk.\mstmtxx{1}, \Ourpk.\mstmtxx{2}, \cdots, \Ourpk.\mstmtxx{t}\| \Oursign.\mstmtxx{1}, \Oursign.\mstmtxx{2}, \cdots, \Oursign.\mstmtxx{t}) \\
& \overset{?}{=} &  \hashH(\Ourpk.\mstmtxx{1}, \cdots, \Ourpk.\mstmtxx{t}\| \Oursign.\mstmtxxp{1}, \cdots, \Oursign.\mstmtxxp{t}).
\end{eqnarray}

Hence, only one (\ie, $\chlgc'$ or $\chlgc$) satisfies the above validations. 
Thus, we have arrived at our contradiction.

\par\noindent\textbf{Collision-Resistance Analysis}. 
Below, we prove the property of collision-resistance using a contradiction.  If $\hashH_2$ is a collision resistant $(\tau$-to-$1$) hash function that every output of $\hashH_2$ has at most $\tau$ preimage in $\G$, we assume that if there happens a collision, then $\Oursign=\hashH_2(\vsk, \hashH_1(\vrfmsg))$ should equal to $\Oursign'=\hashH_2(\vsk, \hashH_1(\vrfmsg'))$, where $\hashh=  \hashH_1(\vrfmsg)$ and $\hashh'= \hashH_1(\vrfmsg')$ for some $\vrfmsg\neq \vrfmsg'$. In this setting, for every $\hashh$, there are at most $\tau$ possible $\hashh'$ values that can cause a collision. Because $\hashh$ and $\hashh'$ are obtained via random oracle queries, then a pair that causes a collision is unlikely to be found after $Q$ queries to $\hashH_1$, as long as the size of $\G$ is larger than $\tau\cdot Q^2/2$.

\par\noindent\textbf{Pseudorandomness Analysis}. Below we analyze the pseudorandomness based on the collision-resistance hash function under the random oracle model. Depending on the \VRF definition, the pseudorandomness definition implies that the pseudorandomness adversary does not know the secret \VRF key $\vsk$, but must distinguish between pairs $(\msgm, \vrfout)$ where $\vrfout$ is the VRF hash output on input $\msgm$, and pairs $(\msgm, r)$ where $r$ is a random value. This adversary knows the public values $\vpk=\hashF(\rkey, \vsk)$, and it can easily compute $\sigma=\hashH_2(\vsk, \hashh)$ and $\hashh=\hashH_1(\msgm)$ for any $\vrfmsg$ if he knows the $\vsk$. However, even $\vpk$ and $\hashh$ are public but $\vsk$ is kept privately, and $\Oursign=\hashH_2(\vsk, \hashH_1(\msgm))$ looks random, thus, the pseudorandomness adversary cannot distinguish $\beta\gets \hashH_3(\Oursign)$ from a randomness distribution because $\beta_0$ is pseudorandom in the range of $\hashH_3$.

\subsection{Post-Quantum VRF Evaluation}

{
\definecolor{Gray}{gray}{0.9}
\definecolor{LightCyan}{rgb}{0.88,1,1}
\newcolumntype{a}{>{\columncolor{Gray}}c}
\newcolumntype{b}{>{\columncolor{white}}c}

\begin{table*}[h!]
\normalsize
\begin{center}
\begin{threeparttable}
    \caption{Performance of Feasibly VRF via \ZKBoo and $\ZKBpp$.}
    \label{table:pqcost}
    \begin{tabular}{l a b a b a b a b a b }
      \toprule 
      \multicolumn{1}{c}{\multirow{3}{*}{\textbf{Scheme}}}
      & \multicolumn{2}{c}{\underline{\textbf{ZKBoo}}}
      & \multicolumn{2}{c}{\underline{\textbf{ZKB++}}}
      \\
      & \multicolumn{2}{c}{\underline{\textbf{$20$-Round}}}
      & \multicolumn{2}{c}{\underline{\textbf{$20$-Round}}}
      \\
      & \textbf{Output.Size}
      & \textbf{Execution.Time}
      & \textbf{Output.Size}
      & \textbf{Execution.Time}\\
      [0.5ex]
      \midrule 
      \multirow{1}{*}{KeyGen}
      & 256~bit  & $<$ 1~ms & 256~bit  & $<$1~ms    \\
      \hdashline
      \multirow{1}{*}{Signing}
      & 256~bit  & $<1$~ms & 256~bit  & $<1$~ms  \\
      \hdashline
      \multirow{1}{*}{Proof}
      & 245920~Byte & 24 ms 
      & 128800~Bytes & 24~ms
      \\
      \hdashline
      \multirow{1}{*}{Verification}
      & \xmark & 16~ms & \xmark & 15~ms  \\

      \bottomrule 
      \multicolumn{1}{c}{\multirow{2}{*}{\textbf{Scheme}}}
      & \multicolumn{2}{c}{\underline{\textbf{$40$-Round}}}
      & \multicolumn{2}{c}{\underline{\textbf{$40$-Round}}}\\
      & \textbf{Output.Size}
      & \textbf{Execution.Time}
      & \textbf{Output.Size}
      & \textbf{Execution.Time}
      \\
      [0.5ex]
      \midrule 
      \multirow{1}{*}{KeyGen}
      & 256~bit & $<1$~ms & 256~bit  & $<1$~ms \\
      \hdashline
      \multirow{1}{*}{Signing}
      & 256~bit  & $<1$~ms & 256~bit  & $<1$~ms \\
      \hdashline
      \multirow{1}{*}{Proof}
      & 491840~Byte & 28~ms
      & 257600~Byte & 29~ms\\
      \hdashline
      \multirow{1}{*}{Verification}
      & \xmark & 18~ms & \xmark & 17~ms \\

      \bottomrule 
      \multicolumn{1}{c}{\multirow{2}{*}{\textbf{Scheme}}}
      & \multicolumn{2}{c}{\underline{\textbf{$60$-Round}}}
      & \multicolumn{2}{c}{\underline{\textbf{$60$-Round}}}\\
      & \textbf{Output.Size}
      & \textbf{Execution.Time}
      & \textbf{Output.Size}
      & \textbf{Execution.Time}\\
      [0.5ex]
      \midrule 
      \multirow{1}{*}{KeyGen}
      & 256~bit & $<1$~ms & 256~bit  & $<1$~ms \\
      \hdashline
      \multirow{1}{*}{Signing}
      & 256~bit  & $<1$~ms & 256~bit  & $<1$~ms \\
      \hdashline
      \multirow{1}{*}{Proof}
      & 737760~Byte & 35~ms
      & 386400~Byte & 36~ms \\
      \hdashline
      \multirow{1}{*}{Verification}
      & \xmark & 23~ms & \xmark & 22~ms  \\

      \bottomrule 
      \multicolumn{1}{c}{\multirow{2}{*}{\textbf{Scheme}}}
      & \multicolumn{2}{c}{\underline{\textbf{$80$-Round}}}
      & \multicolumn{2}{c}{\underline{\textbf{$80$-Round}}}\\
      & \textbf{Output.Size}
      & \textbf{Execution.Time}
      & \textbf{Output.Size}
      & \textbf{Execution.Time}\\
      [0.5ex]
      \midrule 
      \multirow{1}{*}{KeyGen}
      & 256~bit & $<1$~ms & 256~bit  & $<1$~ms \\
      \hdashline
      \multirow{1}{*}{Signing}
      & 256~bit  & $<1$~ms & 256~bit  & $<1$~ms \\
      \hdashline
      \multirow{1}{*}{Proof}
      & 983680~Byte & 42 ms
      & 515200~Byte & 42 ms\\

      \hdashline
      \multirow{1}{*}{Verification}
      & \xmark & 28~ms & \xmark & 25~ms  \\


      \bottomrule 
      \multicolumn{1}{c}{\multirow{2}{*}{\textbf{Scheme}}}
      & \multicolumn{2}{c}{\underline{\textbf{$100$-Round}}}
      & \multicolumn{2}{c}{\underline{\textbf{$100$-Round}}}\\
      & \textbf{Output.Size}
      & \textbf{Execution.Time}
      & \textbf{Output.Size}
      & \textbf{Execution.Time}\\
      [0.5ex]
      \midrule 
      \multirow{1}{*}{KeyGen}
      & 256~bit & $<1$~ms & 256~bit  & $<1$~ms \\
      \hdashline
      \multirow{1}{*}{Signing}
      & 256~bit  & $<1$~ms & 256~bit  & $<1$~ms \\
      \hdashline
      \multirow{1}{*}{Proof}
      & 1249600~Byte & 50 ms 
      & 644000~Byte & 50 ms\\
      \hdashline
      \multirow{1}{*}{Verification}
      &  \xmark & 33~ms  & \xmark & 32~ms  \\
      \bottomrule 
    \end{tabular}
\end{threeparttable}
\end{center}
\vspace{-10pt}
\end{table*}
}

\draft{
\definecolor{Gray}{gray}{0.9}
\definecolor{LightCyan}{rgb}{0.88,1,1}
\newcolumntype{a}{>{\columncolor{Gray}}c}
\newcolumntype{b}{>{\columncolor{white}}c}

\begin{table*}[h!]
\normalsize
\small
\begin{center}
\begin{threeparttable}
    \caption{Performance of Feasibly VRF via \ZKBoo.}
    \label{table:pqcost}
    \begin{tabular}{l a b a b a b a b a b }
      \toprule 
      \multicolumn{1}{c}{\multirow{2}{*}{\textbf{Scheme}}}
      & \multicolumn{2}{c}{\underline{\textbf{$20$-Round}}}
      & \multicolumn{2}{c}{\underline{\textbf{$40$-Round}}}
      & \multicolumn{2}{c}{\underline{\textbf{$60$-Round}}}\\
      & \textbf{Output.Size}
      & \textbf{Execution.Time}
      & \textbf{Output.Size}
      & \textbf{Execution.Time}
      & \textbf{Output.Size}
      & \textbf{Execution.Time}\\
      [0.5ex]
      \midrule 
      \multirow{1}{*}{KeyGen}
      & 256bit  & $<$ 1~ms & 256~bit  & $<$1ms & 256~bit  & $<1$~ms   \\
      \hdashline
      \multirow{1}{*}{Signing}
      & 256~bit  & $<1$~ms & 256~bit  & $<1$~ms & 256~bit  & $<1$~ms \\
      \hdashline
      \multirow{1}{*}{Proof}
      & 249920~Byte & 52 ms & 499840 bytes & 91 ms
      & 749760~Byte & 147 ms \\
      \hdashline
      \multirow{1}{*}{Verification}
      & \xmark & 26 ms & \xmark & 50 ms & \xmark & 72 ms \\

      \bottomrule 
      \multicolumn{1}{c}{\multirow{2}{*}{\textbf{Scheme}}}
      & \multicolumn{2}{c}{\underline{\textbf{$80$-Round}}}
      & \multicolumn{2}{c}{\underline{\textbf{$100$-Round}}}\\
      & \textbf{Execution.Time}
      & \textbf{Output.Size}
      & \textbf{Execution.Time}
      & \textbf{Output.Size}
      \\
      [0.5ex]
      \midrule 
      \multirow{1}{*}{KeyGen}
      & 256~bit & $<1$~ms & 256~bit  & $<1$~ms \\
      \hdashline
      \multirow{1}{*}{Signing}
      & 256~bit  & $<1$~ms & 256~bit  & $<1$~ms \\
      \hdashline
      \multirow{1}{*}{Proof}
      & 999680~Byte & 185 ms
      & 1249600~Byte & 230 ms\\
      \hdashline
      \multirow{1}{*}{Verification}
      & \xmark & 115 ms & \xmark & 139 ms  \\
      \bottomrule 
    \end{tabular}
    \begin{tablenotes}
        \footnotesize
  \item [-]
      \end{tablenotes}
\end{threeparttable}
\end{center}
\vspace{-1em}
\end{table*}
}


To evaluate the performance of \VRF, we are primarily interested in understanding the time consumed in the generation of the proposed signature followed by the hashing time and the total time, including key generation, signing, proof of zero-knowledge, verification, and block formation, when the number of blocks is considered as a part of the blockchain. To evaluate these overheads, we wrote a proof-of-concept blockchain in C++. The developed code is evaluated on Intel\textsuperscript{\textregistered} Core\texttrademark I5-8250U 8th Gen machine with 8GB RAM on an Asus series workstation.


We provide implementations and experimental evaluations of concrete quantum secure \VRF using 
by integrating the post-quantum zero-knowledge proof systems, such as \ZKBoo and \ZKBpp, as shown in Table~\ref{table:pqnizk}. To our knowledge, $\ZKBoo$ and $\ZKBpp$ are one of the hash-based NIZK protocols known to exist in the random oracle model, it is based on the ``MPC-in-the-head'' approach to zero-knowledge, and it can generate (resp. verify) a non-interactive proof for the SHA-1 circuit in approximately 13ms (resp. 5ms), with a proof size of 444KB.
In our realization, to simulate the real-world scene, we used SHA-256 for hashing and was built on top of the $\ZKBoo$ and $\ZKBpp$~\footnote{(\href{https://github.com/Sobuno/ZKBoo}{https://github.com/Sobuno/ZKBoo})}. %
$\ZKBoo$ and $\ZKBpp$ libraries are used to prove knowledge of pre-images of SHA-256.
The following Table~\ref{table:pqcost} gives results for the execution time and output size of key generation, signing, proof, and verification, which is illustrated in terms of time consumed per hash operation, and it presented as an average value after looping 100 times for all the execution modes.

Notably, $\ZKBpp$ is an improved version of $\ZKBoo$ with NIZK proofs that are less than half the size of $\ZKBoo$ proofs, and it shows that this size reduction comes at no extra computational cost. Similarly, in our proof-of-concept realization, the output size of \VRF realized via $\ZKBpp$ is more than halved than \VRF realized by $\ZKB$, not affecting the computational complexity. 
As shown in Table~\ref{table:pqcost}, regarding the execution time of algorithms, such as block formation, key-generation, and signing, each operation is in 1 millisecond, because of which the value of microseconds doesn't make any material difference. In addition, proof generation is the dominant operation, which on average consumes more than 99.0\% of the total execution time. Thus, we use the average time in milliseconds to estimate the running time for these sub-algorithms. These results help to understand the reach of practical observations when such a system is deployed in actual practice. The trade-off between the size and the computational complexity will certainly require configuration considerations, and these results can help in understanding such requirements.

\section{Application~1: Quantum-Safe Decentralized Random Beacon}
\label{sec:beacon}


%
A cryptographic beacon (also known as a randomness beacon) is a service that provides a public source of randomness. The beacon continuously emits new random data (a beacon record) at a regular rate. If everybody agrees that there's no way to predict the following output from the beacon, it can be relied on as a provider of fair random values.
The random values emitted by the beacon can trivially be used for any public lottery,  contract signing, voting protocols, and $\ZKP$ systems, \etc. These applications require random values that cannot be predicted prior to being generated but are made public after generation.
%
Importantly, in the blockchain era,  
many consensus protocols~\cite{FC:DaiPasShi19,Algorand16,DFINITY,EC:DGKR18} involve allocating the creation of block creator, whose selection procedure most often than not requires a method for collective randomness sampling. Obliviously, random beacon
also plays an integral part in the design of new blockchain consensus protocols (\eg, PoS) for a fair and unpredictable distribution of validator responsibilities. The basic reason is that each election begins when public and unbiased random beacon publishes a uniform random number. 
Consequently, how to design the verifiable, bias-resistant, and the unpredictable random number has recently enjoyed increasing attention~\cite{ACNS:CascudoD17,SP:SJKGGKFF17,EPRINT:GLOW20}.
This section gives our random beacon solution via the proposed \VRF while remaining secure in the quantum era. 

%
 %

%



\subsection{Random Beacon Generation}
 
An observation is that if blockchain nodes get different inputs on building blocks, then forks will happen. Thus, to generate uniform randomness, the most straightforward idea is to obtain uniform randomness by hashing past blocks. But the hashes of blocks can be subject to adversarial influence. For example, the random number can be biased in a way so that corrupted nodes are allowed to be selected more often. Thus, no one can guarantee security when this kind of randomness sources with adversarial bias are used for block proposer selection.                                                                        


To address this point, 
Snow White~\cite{FC:DaiPasShi19} proposed a novel ``\textit{two-lookback}'' mechanism to determine present and future randomness depending on its own past blocks, which addresses this kind of a``\textit{randomness-biasing attacks}'' and  ``\textit{adaptive key selection attacks}''  simultaneously, but this approach cannot guarantee perfect uniqueness for randomness as explained earlier. 
%
Ouroboros Paros randomness is secure for block production~\cite{EC:DGKR18} and inherits the incentive
structure of the Ouroboros family. In their approach, all block producers have a \VRF keys, which they register with locked stake, and they depend on a central clock and operate a trusted random beacon using \VRF to board-cast a random number to all participants in each epoch.
In Algorand~\cite{Algorand16}, 
the randomness (\ie, seed) published at round $r$ is determined using \VRF with the seed of the previous round $r-1$, \eg, $\seed_{r}=\hashH(\uSign(\seed_{r-1}) \| r)$, where $\uSign(\seed_{r-1})$ is a deterministic signature $\uSign(\cdot)$ under the leader round $r$, and the initial value of the seed $\seed_0$ bootstraps seed selection. However, the problem is that Algorand does not specify how to initialize $\seed_0$.\footnote{In Algorand, the authors mentioned ``a random number, part of the system description, and thus publicly known.''} In that case, the participant who has set the initial seed $\seed_0$ may have again launched a grinding attack to bias the blockchain consensus protocol.

%
%

Indeed, secret randomness produced by these $\VRF$s can determine when they produce new blocks. A priori, there is a risk that block producers could grind via \VRF keys to bias results. Thus, \VRF inputs must include public randomness created only after the \VRF key. Therefore, during the epoch, the fresh public on-chain randomness is created by hashing together all the \VRF outputs revealed in block creation. In this way, we cycle between private but verifiable randomness and collaborative public randomness.
But an observation is that these kinds of approaches rely on a  trusted party. To avoid reliance on a trusted party, a common approach is to use a mechanism that verifiably allows the distributed computation of an unpredictable and unbiased source of randomness. The distributed $\VRF$s are proposed recently, \eg, \cite{EPRINT:GLOW20} a candidate approach to bypass dependence on the central party. Additionally, how to construct a post-quantum distributed \VRF itself is an exciting research
problem and deserves further deep investigation.



\subsection{Decentralized Random Beacon Generation}
%
%
%

In this section, we resort to the distributed \VRF from symmetric primitives to generate the quantum-safe \textit{decentralized random beacon} (DRB). Very recently, DRSs have recently gained a lot of traction as a key component for leader(s) election in decentralized public ledger technologies. Indeed, as discussed in~\cite{EPRINT:GLOW20}, a conventional DRB provides a way to agree on a randomly chosen leader in a distributed approach for PoS blockchains (\eg,  Dfinity~\cite{DFINITY}, Ethereum 2.0~\cite{Buterin18}, and OmniLedger~\cite{SP:KJGGSF18}), without the need for a coordinator. 
Thus, in this paper, following the research line of DRB, we give a quantum-safe (verifiable) DRB approach. As we know, DRBs are a particular case of verifiable multi-party computation protocol, and they can be straightforwardly obtained from a post-quantum (verifiable) distributed \VRF.

\par\noindent\textbf{Quantum-Safe DRB via Distributed \VRF}.
Our goal is to provide a quantum-safe bias-resistant public randomness in the familiar $(t, n)$-threshold security model in hash-based cryptography and Byzantine consensus protocols. The quantum security can be guaranteed by using hash-based cryptography, as explained in the post-quantum \VRF construction.
Thus, armed with these techniques, in our distributed-\VRF based DRB protocol, we consider the synchronous network where messages are eventually delivered and a Byzantine adversary in $(t,n)$-threshold security model. In this setting, we pick a function with fixed $f+1$ inputs, where $f$ is denoted as the number of dishonest peers, $t=f+1$ is the threshold of verifiable secret sharing, and $n=3f+1$ is denoted as the list of peers that participate in DRB protocol.

%
%
%
%
%
%

\begin{itemize}
\item $(\vsk, \vpk) \gets \DVRF.\DistKG(1^{\secp})$
\begin{itemize}
\item \textbf{Deal phase}. 
\begin{enumerate}
\item The dealer (\ie, each node server) $S_i$ selects a $t$-degree random polynomial $\polyf_i(z)=\sum^t_{k=0} \coeff_{i,k} \cdot z^k =\coeff_{i,0}+\coeff_{i,1} z +\cdots + \coeff_{i,t}z^t \pmod{q}$ over $\Zq$ for $\coeff_{i,k}\in \Zq$.

\item The dealer $S_i$ selects a secret key $\vsk_i:=k_i\gets \Zq$ and sets $\coeff_{i,0}=\polyf_i(0):=\vsk_i$. Next, the dealer computes the corresponding public key {$\vpk_i=\hashF(\rkey, \vsk_i)$} and secret key shares $\ssk_{i,j}:=\polyf_i(j)$ for $1\leq j \leq n$. Then the dealer $S_i$ broadcasts commitments $\cmt_{i,k}\gets\Com(\coeff_{i,k})=\PRG(\seed)\oplus (\coeff_{i,k}\cdot R)$ of each coefficient $\coeff_{i,k}$ for $1 \leq k \leq t$ and the seed $\seed\gets \set{0,1}^{\secp}$ for a pseudorandom generator $\PRG: \set{0,1}^\lambda\to \set{0,1}^{3\secp}$, and a random $R\in \set{0,1}^{3\secp}$. 

\item Output $\vpk_i$, $\set{\ssk_{i,j}}_{j\in [n]}$ and $\set{\cmt_{i,k}}_{k\in[t]}$. 

\end{enumerate}
\item \textbf{Reconstruction phase}. 
\begin{enumerate}
\item Validate 
$\sum_{k=1}^t (\cmt_{i,k}) \cdot ({j^k}) \overset{?}{=} \PRG(\seed)\oplus \big({f_i(j)}\cdot R\big)$ as $\sum_{k=1}^t (\cmt_{i,k}) \cdot ({j^k}) = \PRG(\seed)\oplus \Big(\big(\sum_{k=1}^t \coeff_{i,k}\cdot j^k\big)\cdot R\Big)$.

\item Compute $\revsk_i=\sum_{j\in \textrm{QUAL}} f_j(i)=\sum_{j\in \textrm{QUAL}} \ssk_{i,j}$ and keep it privately. 
\item Output the verification key {$\vvk_{i}=\hashF(\rkey, {\revsk_{i}})$} and the common public key $\vpk^*=\bigoplus_{i\in \textrm{QUAL}} \hashF(\rkey, {\vsk_i})=\hashF({\sum_{i\in \textrm{QUAL}} {\vsk_i} })=\hashF(\rkey, {\vsk^*})$.

\end{enumerate}
\end{itemize}

\item $(\vrfpf, \vrfout)\gets\DVRF.\PEval(\revsk_i, \vvk_i, \vrfmsg)$
\begin{enumerate}
\item Compute $\hashh=\hashH_1(\vrfmsg)$ and $\vrfout_i:=v_i=\hashH_2({\revsk_i}, \hashh)$.
\item Execute the quantum-safe ZKP  \fbox{$\Prove_{\hashH}(\strstmt; \strwits)$} to prove $\vrfout_i:=v_i$ is the correct output given $\revsk_i$ for $\strstmt:=(\hashh, \vrfout_i=\hashH_2({\revsk_i},\hashh), \vvk_i:=\hashF(\rkey, {\revsk_i}))$ and $\strwits:=\revsk_i$. In particular,  two relations $\calL_{\hashF}=\set{(\vvk_i, \rkey; \revsk_i):\vvk_i:=\hashF(\rkey, {\revsk_i})}$ and $\calL_{\hashH_2}=\set{(\vrfout_i, \hashh; \revsk_i): \vrfout_i:=\hashH_2({\revsk_i}, \hashh)}$ are proceeded respectively, as depicted in Fig.\ref{fig:eq-zkboo}. Then return a computation integrity proof $\vrfpf_i=(i, \strchlg_i:=\chlgc_i, \strresp_i:=(\vvk_i.\strresp, \vrfout_i.\strresp))$, where internally the challenge $\chlgc_i$ is involved.
\end{enumerate}

\item $\DVRF.\Comb(\vpk^*, \vvk^*, \vrfmsg, \bigepsilon)$.
\begin{enumerate}
\item Parse list $\bigepsilon=\set{ \vrfpf_{j_1}, \vrfpf_{j_2}, \cdots, \vrfpf_{j_{\abs{\bigepsilon}}} }$ of $\abs{\bigepsilon}\geq t+1$ partial function evaluation candidates originating from $\abs{\bigepsilon}$ different servers, and obtains verification keys $\vvk_{j_1}, \vvk_{j_2}, \cdots, \vvk_{j_{\abs{\bigepsilon}}}$. Notably, $\vvk^*=\sum_{\beta \in [\abs{\bigepsilon}]} \vvk_{j_{\beta}}$.
\item Then, identify an index subset $I=\set{i_1, \cdots, i_{t+1}}$ such that it satisfies the verification of ZKP, \ie, \linebreak
\fbox{$ \ZKP.\Verify_{\hashH}(\strstmt, \vrfpf)$}
  as depicted in Fig.~\ref{fig:eq-v-zkboo} for relations $\calL_{\hashF}$ and $\calL_{\hashH_2}$ respectively. If \linebreak $\ZKP.\Verify_{\hashH}(\strstmt, \vrfpf)=1$
%
%
 (\ie, accept) holds  for every $i\in I$, where $\vrfpf=(i, \strchlg_i, \strresp_i)$, then continue the next step. If no such subset exists, outputs $0$.
\item Reconstruct $\vrfout^*:=v^*(=\hashH_2({\vsk^*}, h))$ for $h=\hashH_1(\vrfmsg)$ can be done from the shares of any qualified set $\textrm{QUAL}$ of participants by calculating the Lagrange interpolation
${v}^*=\sum_{i\in I} (v_i\cdot {\lambda_{0,i,I}}) $, where $v_i=\hashH_2({\revsk_i}, h)$ and $\revsk_i=\sum_{j\in \textrm{QUAL}} f_j(i)=\sum_{j\in \textrm{QUAL}} \ssk_{i,j}$.
\item Output $v^*$ and $\vrfpf^* \gets\set{\vrfpf_i}_{i\in I}$.
\end{enumerate}

\item $\DVRF.\Vrfy(\vpk^*, \vvk^*, \vrfmsg; \vrfpf^*)$
\begin{enumerate}
\item Parse $\vrfpf^*=\set{\vrfpf_i}_{i\in I}$ such that $\abs{I}=t+1$, where $\vrfpf=(i, \strchlg_i, \strresp_i)$ for $i\in I$.
\item Validate $\ZKP.\Verify_{\hashH}(\strstmt, \vrfpf)\overset{?}{=}1$ by revoking the verification $\Verify_{\hashH}(\cdot)$ of the quantum-resistant ZKP. 
 \item Check if $v^* \overset{?}{=} \sum_{i\in I} (v_i\cdot {\lambda_{0,i,I}}) $.
 \end{enumerate}
\end{itemize}


Our proposed unbiasable quantum-safe DRB protocol ensures the properties \textit{quantum security}, \textit{unbiasability}, \textit{unpredictability}, and \textit{availiability}. Additionally, the proposed DRB is practical as it  only depends on hash-based functions with small scale communication overhead. 
%
The post-quantum security is clear because the unbiasable quantum-safe DRB is based on the post-quantum \VRF, so we ignore the detailed analysis. Below, we give a high-level analysis for the other properties. 
\begin{itemize}
\item \textit{Unpredictability} implies that no party (includes the adversarial nodes) could predict (or precompute) anything about the future random output values.
If the participants follow the DRB protocol, the final random number $v^*$ contains $n' \geq  f+1$ secrets. We require there are at most $f$ malicious peers. Thus an adversary cannot obtain the underlying secret before it is revealed or recovered during the specific round. Please refer to~\cite[Theorem~3]{SP:SJKGGKFF17} for details.
 
%
 
 \item \textit{Availability} (also known as liveness) implies that no single party (including the colluding adversary) is allowed to prevent progress. The property is obtained because if the threshold $t = f + 1$ of the verifiable secret-sharing is given, then at least $t$ honest nodes out of the total $2f +1$ positive voters are able to collaborate to recover the secrets. 
 
\item  \textit{Unbiasability} means a random beacon value is statistically indistinguishable from a uniformly random.
Notably, the threshold of verifiable secret sharing $t=f+1$ is given in advance, preventing dishonest participants from recovering the honest secrets. 
If at least $t$ honest participants share messages successfully, then the Byzantine agreement (with at least $2f + 1$ participants) will achieve the validity of these shares as discussed in \textit{availability}. In this case, each honest participant could recover others' secret shares. 
Further, the Byzantine agreement ensures that all honest participants generate a consistent copy of the randomness number. Thus $n'>f$ secrets will be recovered after the barrier point or $n'\leq f$, meaning the protocol fails. 
Thus,  we say the protocol is to prevent the adversary from biasing the random output.
\end{itemize}


\section{Application~2: Quantum-Safe Proof of Stake Consensus Protocol}
\label{sec:VRF}

Obtaining the quantum-safe blockchain straightforwardly from the existing PoW and PoS is non-trivial because the existence of blockchain consensuses, including PoW and PoS, are designed upon the traditional hard-problem assumptions (\ie, cryptographic puzzles) without the property of quantum resistance. 
In essence, PoS consensus protocols (\eg, Snow White~\cite{FC:DaiPasShi19}, DFINITY~\cite{DFINITY}, and Algorand~\cite{Algorand16} \etc) are similar to PoW protocols to some extent, because they are from proof systems, and the participants elect the creator who can create the next block.
In particular, regarding chain-based leader election in PoS consensus, Ouroboros~\cite{C:KRDO17} replaces energy-consuming cryptographic puzzles by finding a solution (\ie, $\sigma_i=\Sign_{\sk_i}(\state_i, \data_i, \slot_j))$ to satisfy the verification requirement of $\Vrfy_{\pk_i}(\sigma_i, (\state_i, \data_i, \slot_j))=1$ for the state $\state_i=\hashH(\blockB_{i-1})$, where the signature is the existential unforgeability under chosen message attack resistance scheme. Similarly, the lottery-based consensus mechanism (\eg, Algorand~\cite{Algorand16}) is based on a fast Byzantine agreement protocol, and the agreement is not performed between all users in the network. Instead, it is confined to a small randomly chosen committee of users for each round. Algorand pioneered the use of VRF for secret encryption and lottery to run a consensus agreement with the election committee. This enables the Algorand blockchain to achieve the scale and performance required to support millions of users.
The core election process in Algorand~\cite{Algorand16} can be expressed as $\VRF(\hashh(\blockB_i), \strround, \pk, \sigma)< \targetT$, where the signature $\sigma$ is a result of a (valid) payment pay relative to the stake (or the transfer amounts of money units) $a$. 
In that case, we explore the utility of the proposed quantum-secure \VRF in the lottery-based Algorand. 

To obtain the quantum-safe PoS, we trim the quantum-secure \VRF inequality by denoting the difficulty value dynamically as the product of the stake owned by each elector and the target value fixed by the system. Further, following the methodology of Algorand, every user performs secret self-selection based on his/her \VRF secret key for each block. If the quantum-secure \VRF value that the user obtains is less than some threshold, then the user is selected to serve on the committee to perform an agreement for the block.
{Then, in each round, each node needs to check to evaluate a quantum-secure \VRF to check whether they have been sampled as part of that round's committee. The number of committee members is binomial (\ie, depending on the stake distribution), so the amount of work that needs to be done to verify the other committee members' messages is also binomial.}

\par\noindent\textbf{Evaluation of Transaction}.
We implement a proof of concept of our scheme to compare against the Algorand VRF. Our evaluation indicates that our
construction does not introduce any prohibitive overheads, further, the proposed constructions are practiced even today. For instance, the transmission of
the overhead introduced per every block is an additional 824KB which accounts for around 8\%
of the 10MB Algorand block size. 
To evaluate the block formation for quantum-secure PoS consensus, when the number of blocks is considered as a part of the blockchain. The prover program takes as input the secret key and simulated stakes, then the prover program will generate the proof of computation of $\hashH(\vsk)$ and $\hashH(\vsk\|\text{stakes})$ using $\ZKBoo$ or $\ZKBpp$. In our proof of concept implementation, proof generation using $\ZKBpp$ takes an additional 61 milliseconds overhead when comparing with the Algorand VRF.
Meanwhile, the verifier will check the proof to be sure that the prover knows the pre-image of $\vsk$ and owns stakes. Notably, the verification step is the most time-consuming part.
In PoS-blockchain, the results will vary as per the number of rounds of $\ZKBoo$ (or $\ZKBpp$) used in each round of every epoch in the underlying protocol. This can be understood in Table~\ref{table:pqcost}, which shows that with additional rounds, the security will increase for a system using either of the approaches. There is a trade-off that for every epoch in PoS, the rounds of $\ZKBoo$ (or $\ZKBpp$) will also cause execution overheads. However, given the advantages of security and not so tighter impact on the performance, these solutions can be considered for implementing quantum-safe blockchains.


We emphasize that designing a practical post-quantum \VRF for quantum-safe PoS consensus is an important urgent open problem.
First of all, blockchain systems should be long-lived by design. From their current deployments, we know that any (even purely technical) protocol-level change is challenging to be introduced in practice and may result in a platform fork, where part of the community
``believes'' in the old system (without the change) and other parts in the new
system (with the change applied).\footnote{In fact, such forks happened
to mainstream platforms like Bitcoin or Ethereum.}
Secondly, due to their potential and promises, the security of blockchain
systems should be treated critically.
Even though one could argue that quantum computing is not a short-term threat,
we stress that due to deployability and governance of these systems it may be
challenging to update them in the future.


\draft{

 $\ZKBoo$ and $\zkstark$ library, respectively.

 which

Further,

zk-STARK is a zero-knowledge scalable and transparent argument of knowledge and it is considered to be resistant to advances in quantum computing. Notice the key differences to zk-SNARKs (\ie, zero-knowledge succinct non-interactive argument of knowledge), namely scalability and transparency.
In a nutshell, first and foremost, zk-STARKs have solved the trusted setup problem, because of this, they are said to be \textit{transparent}. They completely remove the need for multiple parties to create the private key needed for the string. Instead, everything needed to generate the proofs is public and the proofs are generated from random numbers. zk-STARKs actually remove the requirement in zk-SNARKs for asymmetric cryptography and instead use the hash functions. Beyond this, they should have a longer shelf life in terms of their cryptographic resilience than zk-SNARKs.
}

%
%
%
%
%
%
%
%
%
%
%
%
%
%
%
%
%
%
%
%
%
%
%
%
%
%
%
%
%

\section{Conclusion}
\label{sec:concl}

The focus of this work is to explore the probability of how to design a post-quantum random beacon and PoS consensus layer. Therefore, the crux of fulfilling the main goal is turning into how to construct a practical post-quantum \VRF. Inspired by the relationship between the unique signature and \VRF, we started with the quantum-resistant hash-based signature and integrated the signature with the post-quantum ZKP system to obtain a post-quantum \VRF.
%
More specifically, a quantum-resistant PoS consensus framework is instantiated with our proposed post-quantum \VRF by combining the hash functions and quantum-secure ZKP from symmetric primitives. Furthermore, we rigorously analyzed and proved the security of our construction. Additionally, we implemented a proof of concept of the system supported by the \ZKBoo and \ZKBpp, and our conducted experiments indicate that the scheme is deployable even as for today. 

Notably, defining the application layer and its security is out of scope for this work.
However, in the future, we would like to extend our system by a post-quantum application layer, where user transactions are also guaranteed quantum-resistant without introducing prohibitive overheads. Further, we just provided a proof-of-concept realization. The existing post-quantum ZKP system can be used in our \VRF construction, such as Ligero and \zkstark. We leave these works in the future.

\ifCLASSOPTIONcaptionsoff
  \newpage
\fi


{
{
\def\shortbib{0}
{
\begin{spacing}{1}
\normalem
  \bibliographystyle{IEEEtran}
 \bibliography{crypto/BlockChain}
\end{spacing}
}}


\end{document}